# The Re-Encoding Transformation in Algebraic List-Decoding of Reed-Solomon Codes

Ralf Koetter, *Fellow, IEEE*, Jun Ma, *Senior Member, IEEE*, and Alexander Vardy, *Fellow, IEEE*

*Abstract* — The main computational steps in algebraic soft-decoding, as well as Sudan-type list-decoding, of Reed-Solomon codes are bivariate polynomial interpolation and factorization. We introduce a computational technique, based upon re-encoding and coordinate transformation, that significantly reduces the complexity of the bivariate interpolation procedure. This re-encoding and coordinate transformation converts the original interpolation problem into another *reduced interpolation problem*, which is orders of magnitude smaller than the original one. A rigorous proof is presented to show that the two interpolation problems are indeed equivalent. An efficient factorization procedure that applies directly to the reduced interpolation problem is also given.

## I. Introduction

REED-Solomon codes are still widely used in digital communications and data storage. Conventional hard-decision Reed-Solomon decoders correct up to $(n-k)/2$ symbol errors for an $(n,k)$ Reed-Solomon code. During the past decade, several breakthroughs in improving the error-correction capability of Reed-Solomon decoders have been achieved. First, Sudan [20] proved that list-decoding of Reed-Solomon codes can be viewed as a bivariate interpolation problem, thereby correcting more errors than previously thought possible. Specifically, for an $(n,k)$ Reed-Solomon code, Sudan's algorithm [20] produces all codewords whose distance to the received hard-decision vector is at most $n - \sqrt{2kn}$. The second major step was taken by Guruswami and Sudan [8]. Their work showed that one can correct even more errors by interpolating through each point not once, but $m$ times, where $m$ is an integer parameter. For $m \to \infty$, the list-decoding algorithm of [8] corrects up to $n - \sqrt{nk}$ errors. These results were later extended by Koetter and Vardy [12], who showed how the interpolation multiplicities in the algorithm of [8] should be chosen to achieve algebraic soft-decision decoding of Reed-Solomon codes.

Both list-decoding and algebraic soft-decision decoding rely upon interpolation and factorization of bivariate polynomials. Such interpolation and factorization is much more computationally intensive than the Berlekamp-Massey algorithm used in conventional hard-decision decoding. Various efficient algorithms for bivariate interpolation and factorization have been proposed by Alekhnovich [2], Augot-Pecquet [3], Feng-Giraud [4], Gao-Shokrollahi [5], Koetter [9], Lee-O'Sullivan [13], Nielsen-Høholdt [16], Olshevsky-Shokrollahi [17], Roth-Ruckenstein [18], Wu-Siegel [24], among others. While polynomial-time, these algorithms fall short of making the required computation feasible in practical applications that usually involve long high-rate Reed-Solomon codes.

In this paper, we present a series of transformations that drastically reduce the space and time complexity of the bivariate interpolation process, by a factor of at least $n^2/(n-k)^2$. These transformations were first reported in the conference papers [6], [10], and [11]. Later, the resulting re-encoding and coordinate transformation procedures were used, without proof, in [1,7,14] and other papers. Our main goal herein is to provide a streamlined formulation of this transformation process, and of the corresponding factorization procedure, including rigorous proofs and detailed examples.

The rest of this paper is organized as follows. In Section II, we introduce the necessary definitions and establish (in Theorem 1) a basic property of the birational isomorphism that underlies the coordinate transformation process. The bivariate interpolation problem encountered in list-decoding and in algebraic soft-decoding of Reed-Solomon codes is briefly reviewed in Section III. We also review in Section III the Gröbner-basis bivariate interpolation algorithm due to Koetter [9]. The re-encoding and coordinate transformation complexity-reduction method is described in detail in Section IV. This section also contains a rigorous proof of the validity of this method. In Section V, we show how the last decoding step, namely bivariate polynomial factorization, should be carried out in order to take full advantage of the reduced interpolation procedure of Section IV. We conclude with a brief summary in Section VI.

## II. Definitions and Preliminaries

Let $\mathbb{F}_q$ be the finite field with $q$ elements. The ring of polynomials over $\mathbb{F}_q$ is denoted $\mathbb{F}_q[X]$. Reed-Solomon codes are obtained by evaluating certain polynomials in $\mathbb{F}_q[X]$ on a set of points $\mathcal{D} \subseteq \mathbb{F}_q$. Specifically, the Reed-Solomon code $\mathbb{C}_q(n,k)$ of length $n$ and dimension $k$ is defined as follows:

$$\mathbb{C}_q(n,k) \stackrel{\text{def}}{=} \Big\{ (f(\xi_1), \ldots, f(\xi_n)) \ : \ \xi_1, \xi_2, \ldots, \xi_n \in \mathcal{D}, \\ f(X) \in \mathbb{F}_q[X], \ \deg f(X) < k \Big\}$$

The set $\mathcal{D}$ is usually taken as $\mathbb{F}_q$ or as $\mathbb{F}_q^*$, where $\mathbb{F}_q^*$ is the set of all the nonzero elements of $\mathbb{F}_q$. Unless stated otherwise, we shall assume that $\mathcal{D} = \mathbb{F}_q^*$, so that $n = q-1$. As in [8,12,18], we define the weighted degree of a polynomial as follows.

The material in this paper was presented in part at the Information Theory Workshop, Paris, France, April 2003, and at the IEEE International Symposium on Information Theory, Yokohama, Japan, July 2003.

Ralf Koetter was with the Institute for Communications Engineering, Technische Universität München, D-80290 Munich, Germany.

Jun Ma is with Qualcomm, 5775 Morehouse Drive, San Diego, CA 92121, U.S.A. (e-mail: jun.jma@gmail.com).

Alexander Vardy is with the Department of Electrical and Computer Engineering, the Department of Computer Science and Engineering, and the Department of Mathematics, all at the University of California San Diego, La Jolla, CA 92093–0407, U.S.A. (e-mail: avardy@ucsd.edu).

**Definition 1.** *Let $\mathcal{A}(X,Y) = \sum_{i=0}^{\infty}\sum_{j=0}^{\infty} a_{i,j} X^i Y^j$ be a bivariate polynomial and let $w_X, w_Y$ be arbitrary real numbers. Then the $(w_X, w_Y)$-**weighted degree** of $\mathcal{A}(X,Y)$ is defined as the maximum over all real numbers $i w_X + j w_Y$ such that $a_{i,j} \neq 0$.*

For reasons that will become clear later, we do not restrict the definition of weighted degree to the usual case [8,12,18] where $w_X, w_Y$ are nonnegative integers. Thus the weighted degree of a polynomial $\mathcal{A}(X,Y)$ can assume negative values.

Let $K_{\alpha,\beta}$ denote the ring of rational functions in $\mathbb{F}_q(X,Y)$ without poles at the point $(\alpha,\beta) \in \mathbb{F}_q \times \mathbb{F}_q$. A rational function $\mathcal{A}(X,Y) \in K_{\alpha,\beta}$ always has a power-series expansion in basis functions of type $(X-\alpha)^i(Y-\beta)^j$. Thus we can write

$$\mathcal{A}(X,Y) = \sum_{i=0}^{\infty}\sum_{j=0}^{\infty} a_{i,j}(X-\alpha)^i(Y-\beta)^j \qquad (1)$$

and define

$$\mathbf{coef}\big(\mathcal{A}(X+\alpha, Y+\beta); X^i Y^j\big) \stackrel{\text{def}}{=} a_{i,j} \qquad (2)$$

**Definition 2.** *The function $\mathcal{A}(X,Y)$ is said to **pass through the point** $(\alpha,\beta)$ **with multiplicity** $m$ if $a_{i,j} = 0$ for all $i + j < m$ in the power-series expansion (1). Further, the **multiplicity function** $\mu_{\alpha,\beta}: K_{\alpha,\beta} \to \mathbb{N}$ is defined as follows:*

$$\mu_{\alpha,\beta}\big(\mathcal{A}(X,Y)\big) \stackrel{\text{def}}{=} \max\{m \in \mathbb{N} \,:\, a_{i,j} = 0 \; \forall\, i+j < m\}$$

The following properties of the multiplicity function follow from the well-known fact that $\mu_{\alpha,\beta}$ is an (exponential) valuation. Given any $\mathcal{A}(X,Y)$ and $\mathcal{B}(X,Y)$ in $K_{\alpha,\beta}$, we have

$$\mu_{\alpha,\beta}\big(\mathcal{A}(X,Y)\big) = \mu_{0,0}\big(\mathcal{A}(X+\alpha, Y+\beta)\big) \qquad (3)$$
$$\mu_{\alpha,\beta}(\mathcal{AB}) = \mu_{\alpha,\beta}(\mathcal{A}) + \mu_{\alpha,\beta}(\mathcal{B}) \qquad (4)$$
$$\mu_{\alpha,\beta}(\mathcal{A}+\mathcal{B}) \geq \min\{\mu_{\alpha,\beta}(\mathcal{A}), \mu_{\alpha,\beta}(\mathcal{B})\} \qquad (5)$$

Assuming that $\mathcal{A}(X,Y)/\mathcal{B}(X,Y) \in K_{\alpha,\beta}$, we also have

$$\mu_{\alpha,\beta}\left(\frac{\mathcal{A}}{\mathcal{B}}\right) = \mu_{\alpha,\beta}(\mathcal{A}) - \mu_{\alpha,\beta}(\mathcal{B}) \qquad (6)$$

**Definition 3.** *Fix a polynomial $g(X)$ over $\mathbb{F}_q$, and let $\mathfrak{Z}$ be the set of all $\alpha \in \mathbb{F}_q$ such that $g(\alpha) = 0$. We define the following pair of mappings between points in $(\mathbb{F}_q - \mathfrak{Z}) \times \mathbb{F}_q$:*

$$\varphi_g: (x,y) \mapsto \left(x, \frac{y}{g(x)}\right) \qquad (7)$$
$$\varphi_g^{-1}: (x,y) \mapsto (x, y g(x)) \qquad (8)$$

It is easy to see that $\varphi_g$ and $\varphi_g^{-1}$ are birational isomorphisms, and $\varphi_g^{-1}$ is indeed the inverse of $\varphi_g$.

**Definition 4.** *Fix a polynomial $g(X)$ over $\mathbb{F}_q$. Then the maps $\Phi_g$ and $\Phi_g^{-1}$ from $\mathbb{F}_q(X,Y)$ to itself are defined as follows:*

$$\Phi_g: \sum_{j=0}^{\infty}\sum_{i=0}^{\infty} q_{i,j} X^i Y^j \;\mapsto\; \sum_{j=0}^{\infty}\sum_{i=0}^{\infty} q_{i,j} X^i g(X)^j Y^j$$
$$\Phi_g^{-1}: \sum_{j=0}^{\infty}\sum_{i=0}^{\infty} q_{i,j} X^i Y^j \;\mapsto\; \sum_{j=0}^{\infty} \frac{\sum_{i=0}^{\infty} q_{i,j} X^i}{g(X)^j} Y^j \qquad (9)$$

Thus $\Phi_g$, respectively $\Phi_g^{-1}$, maps a rational function $\mathcal{A}(X,Y)$ into $\mathcal{A}(X, Y g(X))$, respectively $\mathcal{A}(X, Y/g(X))$.

The following theorem establishes an important property of the map $\Phi_g$ that will be crucial for our results in Section IV.

**Theorem 1.** *Let $(\alpha,\beta) \in (\mathbb{F}_q - \mathfrak{Z}) \times \mathbb{F}_q$, and let $\mathcal{A}(X,Y) \in K_{\alpha,\beta}$. Define $(\alpha,\gamma) = \varphi_g(\alpha,\beta)$ and $\mathcal{B}(X,Y) = \Phi_g(\mathcal{A}(X,Y))$. Then*

$$\mu_{\alpha,\beta}\big(\mathcal{A}(X,Y)\big) = \mu_{\alpha,\gamma}\big(\mathcal{B}(X,Y)\big)$$

*Proof.* Since, by assumption, $\mathcal{A}(X,Y)$ does not have a pole at $(\alpha,\beta)$, and $\mathcal{B}(\alpha,\gamma) = \mathcal{A}(\alpha, \gamma g(\alpha)) = \mathcal{A}(\alpha,\beta)$, we see that $\mathcal{B}(X,Y)$ does not have a pole at $(\alpha,\gamma)$. Therefore, we can expand $\mathcal{B}(X,Y)$ in basis functions $(X-\alpha)^i(Y-\gamma)^j$ as follows:

$$\mathcal{B}(X,Y) = \sum_{i=0}^{\infty}\sum_{j=0}^{\infty} b_{i,j}(X-\alpha)^i(Y-\gamma)^j \qquad (10)$$

With the coefficients $b_{i,j}$ determined by (10), our definition of the multiplicity function implies that

$$\mu_{\alpha,\gamma}\big(\mathcal{B}(X,Y)\big) = \min_{b_{i,j}\neq 0}\{i+j\}$$

Since the maps $\Phi_g$ and $\Phi_g^{-1}$ in (9) are inverses of each other, we can express $\mathcal{A}(X,Y)$ as follows:

$$\mathcal{A}(X,Y) = \Phi_g^{-1}\big(\mathcal{B}(X,Y)\big) = \mathcal{B}\left(X, \frac{Y}{g(X)}\right) \qquad (11)$$
$$= \sum_{i=0}^{\infty}\sum_{j=0}^{\infty} b_{i,j}(X-\alpha)^i\left(\frac{Y}{g(X)} - \gamma\right)^j \qquad (12)$$
$$= \sum_{i=0}^{\infty}\sum_{j=0}^{\infty} b_{i,j}(X-\alpha)^i\left(\frac{Y - \gamma g(X)}{g(X)}\right)^j \qquad (13)$$

Dividing $X-\alpha$ into $g(X)$ with remainder $g(\alpha)$, we can write $g(X) = (X-\alpha)h(X) + g(\alpha)$ and

$$\gamma g(X+\alpha) = \gamma X h(X+\alpha) + \gamma g(\alpha) = \beta + \gamma X h(X+\alpha)$$

for a nonzero polynomial $h(X)$. With this, it follows from (3), (5), and (13) that

$$\mu_{\alpha,\beta}(\mathcal{A}) \geq \min_{b_{i,j}\neq 0}\left\{\mu_{\alpha,\beta}\left((X-\alpha)^i\left(\frac{Y-\gamma g(X)}{g(X)}\right)^j\right)\right\}$$
$$= \min_{b_{i,j}\neq 0}\left\{\mu_{0,0}\left(X^i\left(\frac{Y+\beta-\gamma g(X+\alpha)}{g(X+\alpha)}\right)^j\right)\right\}$$
$$= \min_{b_{i,j}\neq 0}\left\{\mu_{0,0}\left(X^i\left(\frac{Y-\gamma X h(X+\alpha)}{g(X+\alpha)}\right)^j\right)\right\}$$

Since $g(\alpha) \neq 0$, we can use (4) and (6) to further simplify the right-hand side above as follows:

$$\mu_{0,0}\left(X^i\left(\frac{Y-\gamma X h(X+\alpha)}{g(X+\alpha)}\right)^j\right) =$$
$$i\mu_{0,0}(X) + j\mu_{0,0}\big(Y-\gamma X h(X+\alpha)\big) - j\mu_{0,0}\big(g(X+\alpha)\big)$$

It is clear that $\mu_{0,0}(X) = \mu_{0,0}(Y-\gamma X h(X+\alpha)) = 1$, while $\mu_{0,0}(g(X+\alpha)) = 0$. Hence, we finally conclude that

$$\mu_{\alpha,\beta}\big(\mathcal{A}(X,Y)\big) \geq \min_{b_{i,j}\neq 0}\{i+j\} = \mu_{\alpha,\gamma}\big(\mathcal{B}(X,Y)\big) \qquad (14)$$

To establish the inequality in the other direction a similar argument can be used. First expand $\mathcal{A}(X,Y)$ in basis functions of type $(X-\alpha)^i(Y-\beta)^j$ in order to define the coefficients $a_{i,j}$ such that $\mu_{\alpha,\beta}(\mathcal{A}(X,Y)) = \min_{a_{i,j}\neq 0}\{i+j\}$. Then write

$$\begin{aligned}\mathcal{B}(X,Y) &= \Phi(\mathcal{A}(X,Y)) = \mathcal{A}(X,Yg(X)) \\ &= \sum_{i=0}^{\infty}\sum_{j=0}^{\infty} a_{i,j}(X-\alpha)^i(Yg(X)-\beta)^j\end{aligned}$$

to conclude that

$$\begin{aligned}\mu_{\alpha,\gamma}(\mathcal{B}) &\geqslant \min_{a_{i,j}\neq 0}\left\{\mu_{\alpha,\gamma}\big((X-\alpha)^i(Yg(X)-\beta)^j\big)\right\} \\ &= \min_{a_{i,j}\neq 0}\left\{\mu_{0,0}\big(X^i((Y+\gamma)g(X+\alpha)-\beta)^j\big)\right\}\end{aligned}$$

But $(Y+\gamma)g(X+\alpha) - \beta = Yg(X+\alpha) + \gamma Xh(X+\alpha)$ and it is easy to see that $\mu_{0,0}(Yg(X+\alpha) + \gamma Xh(X+\alpha)) = 1$. This shows that $\mu_{\alpha,\gamma}(\mathcal{B}(X,Y)) \geqslant \min_{a_{i,j}\neq 0}\{i+j\}$ and, in conjunction with (14), completes the proof of the theorem. ∎

## III. THE INTERPOLATION PROBLEM

Our interest in the foregoing definitions and results is motivated by the fact that, as a consequence of Bezout's theorem, two polynomials $\mathcal{A}(X,Y)$ and $\mathcal{B}(X,Y)$ cannot both pass with high multiplicity through an arbitrary large number of points without having a common factor. In particular, the polynomial $Y - f(X)$, with $\deg f(X) < k$, passes through the $n$ points $(\xi_1, c_1), (\xi_2, c_2), \ldots, (\xi_n, c_n)$, where $c_i = f(\xi_i)$ may be thought of as the $n$ transmitted symbols. Then Bezout's theorem implies that any nonzero polynomial $\mathcal{Q}(X,Y)$ such that

$$\sum_{i=1}^{n}\mu_{\xi_i,c_i}(\mathcal{Q}) > \deg_{1,k-1}\mathcal{Q}(X,Y)$$

is divisible by $Y - f(X)$. This observation leads to the interpolation-based decoding algorithms of [8,20] and [12]. The central idea of these decoding algorithms is to construct a polynomial $\mathcal{Q}(X,Y)$ that passes through a prescribed set of points

$$\mathcal{P} \stackrel{\text{def}}{=} \big\{(x_1,y_1),(x_2,y_2),\ldots,(x_s,y_s)\big\} \quad (15)$$

where $x_1, x_2, \ldots, x_s \in \mathcal{D}$ and $y_1, y_2, \ldots, y_s \in \mathbb{F}_q$, with the prescribed multiplicities $m_{x_1,y_1}, m_{x_2,y_2}, \ldots, m_{x_s,y_s}$, which are positive integers. If these points and multiplicities agree "sufficiently well" with the $n$ points $(\xi_i, c_i)$ that determine the transmitted codeword, then divisibility of $\mathcal{Q}(X,Y)$ by $Y - f(X)$ is guaranteed. The reader is referred to [12] for more details.

In all cases, a key part of the decoding algorithm consists of solving the following interpolation problem.

**Definition 5 (Interpolation problem).** *Given a set of $s$ points $\mathcal{P} = \big\{(x_1,y_1),(x_2,y_2),\ldots,(x_s,y_s)\big\}$ and a set of $s$ multiplicities $M = \{m_{x_1,y_1}, m_{x_2,y_2}, \ldots, m_{x_s,y_s}\}$, the interpolation problem consists of computing a polynomial $\mathcal{Q}(X,Y) \not\equiv 0$ such that*

$$\mu_{x_i,y_i}(\mathcal{Q}(X,Y)) \geqslant m_{x_i,y_i} \quad \text{for all } (x_i,y_i) \in \mathcal{P} \quad (16)$$

*and $\deg_{1,k-1}\mathcal{Q}(X,Y)$ is minimal among all bivariate polynomials that satisfy the interpolation constraints (16).*

We shall refer to this interpolation problem as $\mathbf{IP}_{1,k-1}(\mathcal{P}, M)$, and say that $\mathcal{Q}(X,Y)$ is a ***solution to*** $\mathbf{IP}_{1,k-1}(\mathcal{P}, M)$. Observe that $x_i$ and $x_j$ in the set $\mathcal{P} = \{(x_1,y_1),(x_2,y_2),\ldots,(x_s,y_s)\}$ do not have to be distinct, all we require is that they belong to $\mathcal{D}$. In fact, in soft-decision decoding, we often interpolate through different points having the same X-coordinate [12].

By definition (see Definition 2), requiring that a polynomial $\mathcal{Q}(X,Y)$ passes through a given point with multiplicity $m$ imposes $\frac{1}{2}m(m+1)$ linear constraints on the vector space of polynomials in two variables. Hence, solving the interpolation problem $\mathbf{IP}_{1,k-1}(\mathcal{P}, M)$ is tantamount to solving a system of

$$N(M) \stackrel{\text{def}}{=} \frac{1}{2}\sum_{i=1}^{s} m_{x_i,y_i}(m_{x_i,y_i}+1) \quad (17)$$

homogeneous linear equations. As shown in [8], [12], and other papers, there are precisely

$$\chi_{1,k-1}(\delta) \stackrel{\text{def}}{=} \left\lceil \frac{\delta+1}{k-1} \right\rceil \left(\delta - \frac{k-1}{2}\left\lfloor \frac{\delta}{k-1} \right\rfloor + 1\right) \quad (18)$$

monomials $X^iY^j$ with $i + (k-1)j \leqslant \delta$. Hence, choosing $\delta$ to be large enough will guarantee a solution to $\mathbf{IP}_{1,k-1}(\mathcal{P}, M)$. Let $\delta^*$ be the least integer such that $\chi_{1,k-1}(\delta^*) > N(M)$. Then the weighted-degree of a solution $\mathcal{Q}(X,Y)$ to $\mathbf{IP}_{1,k-1}(\mathcal{P}, M)$ satisfies $\deg_{1,k-1}\mathcal{Q}(X,Y) \leqslant \delta^*$, and its $Y$-degree is at most

$$r \stackrel{\text{def}}{=} \left\lfloor \frac{\delta^*}{k-1} \right\rfloor \quad (19)$$

In principle, $\mathbf{IP}_{1,k-1}(\mathcal{P}, M)$ is a simple linear problem that can be solved in a number of ways. Numerous algorithms for this purpose have been proposed in [1,2,4,6,7,9,13,16–18,22]. In what follows, we briefly review the interpolation algorithm of Koetter [9], which is widely recognized as one of the most suitable for implementation in practice. VLSI architecture for this algorithm has been developed in [1,6,7] and other papers.

In order to describe Koetter's interpolation algorithm [9], we first need to extend the notion of $(1,k-1)$-weighted degree to a monomial order. Explicitly, we say that $X^aY^b \prec_k X^iY^j$ iff

$$\begin{aligned} a + (k-1)b &< i + (k-1)j \\ &\textbf{or} \\ a + (k-1)b &= i + (k-1)j \quad \text{and} \quad b < j \end{aligned} \quad (20)$$

Every polynomial in $\mathbb{F}_q[X,Y]$ now has a well-defined leading term under $\prec_k$, and we can impose a total pre-order $\prec_k$ on polynomials in $\mathbb{F}_q[X,Y]$ by comparing their leading terms.

Notably, Koetter's algorithm computes much more than just a solution to the interpolation problem $\mathbf{IP}_{1,k-1}(\mathcal{P}, M)$. It produces a <u>set</u> of polynomials, which forms a Gröbner basis for the ideal of all polynomials over $\mathbb{F}_q$ that satisfy the interpolation constraints defined by $\mathcal{P}$ and $M$. This Gröbner basis $\mathscr{G}$ consists of $r+1$ polynomials: $\mathscr{G} = \{\mathcal{G}_0, \mathcal{G}_1, \ldots, \mathcal{G}_r\}$, where $r$ is given by (19). The computation is initialized by setting

$$\mathscr{G}^{(0)} := \{1, Y, Y^2, \ldots, Y^r\} \quad (21)$$

Upon initialization, Koetter's algorithm goes through $N(M)$ iterations, imposing each of the $N(M)$ linear constraints in (17) one-by-one. For example, suppose that at iteration $i$ of the algorithm, we are dealing with the constraint

$$\mathbf{coef}\big(\mathcal{Q}(X+x, Y+y); X^aY^b\big) = 0$$

for some $(x,y) \in \mathcal{P}$ and some $a + b < m_{x,y}$. Then, given the set $\mathscr{G}^{(i-1)} = \{\mathcal{G}_0^{(i-1)}, \mathcal{G}_1^{(i-1)}, \ldots, \mathcal{G}_r^{(i-1)}\}$ computed after the first $i-1$ iterations, the $i$-th iteration consists of the following.

**UpdateBasis**$(\mathcal{G}; (x,y); a, b)$

**1** For all $j = 0, 1, \ldots, r$, compute the *discrepancy* of $\mathcal{G}_j^{(i-1)}$ which is given by:
$$\Delta_j \stackrel{\text{def}}{=} \textbf{coef}\big(\mathcal{G}_j^{(i-1)}(X+x, Y+y); X^a Y^b\big)$$
If $\Delta_j = 0$ for all $j = 0, 1, \ldots, r$, set $\mathcal{G}^{(i)} := \mathcal{G}^{(i-1)}$ and stop.

**2** Among $\mathcal{G}_0^{(i-1)}, \mathcal{G}_1^{(i-1)}, \ldots, \mathcal{G}_r^{(i-1)}$, find the least with respect to $\prec_k$ polynomial such that its discrepancy is nonzero. Let $\mathcal{G}_t^{(i-1)}(X, Y)$ denote this *pivot polynomial*, so $\Delta_t \neq 0$.

**3** For all $j = 0, 1, \ldots, r$, except $j = t$, compute
$$\mathcal{G}_j^{(i)}(X, Y) := \mathcal{G}_j^{(i-1)}(X, Y) - \frac{\Delta_j}{\Delta_t} \mathcal{G}_t^{(i-1)}(X, Y)$$
Then update the pivot polynomial $\mathcal{G}_t^{(i-1)}(X, Y)$, namely set
$$\mathcal{G}_t^{(i)}(X, Y) := (X - x)\, \mathcal{G}_t^{(i-1)}(X, Y)$$

It can be shown [22] that for all $i$, the $Y$-degree of the leading monomial of $\mathcal{G}_j^{(i)}(X, Y)$ is exactly $j$. Along with (20), this implies that the polynomials $\mathcal{G}_0^{(i)}, \mathcal{G}_1^{(i)}, \ldots, \mathcal{G}_r^{(i)}$ always have distinct orders with respect to $\prec_k$ and guarantees that the choice of the pivot polynomial in Step **2** is unique.

Using the **UpdateBasis**$(\mathcal{G}; (x,y); a, b)$ procedure above, the entire interpolation algorithm can be formulated as follows.

**Koetter's Interpolation Algorithm**

**Input:** A set of points $\mathcal{P} = \{(x_1, y_1), (x_2, y_2), \ldots, (x_s, y_s)\}$ along with their multiplicities $m_{x_1, y_1}, m_{x_2, y_2}, \ldots, m_{x_s, y_s}$.

**Initialization:** Set $\mathcal{G} := \{1, Y, Y^2, \ldots, Y^r\}$.

**Iterations:** For all $(x, y) \in \mathcal{P}$, do the following: for $a := 0$ to $m_{x,y} - 1$, then for $b := 0$ to $m_{x,y} - a - 1$, update $\mathcal{G}$ using the procedure **UpdateBasis**$(\mathcal{G}; (x,y); a, b)$.

**Output:** Return the least with respect to $\prec_k$ polynomial in $\mathcal{G}$.

It is not difficult to see that the number of additions and multiplications in $\mathbb{F}_q$ required to solve $\mathbf{IP}_{1,k-1}(\mathcal{P}, M)$ using this algorithm is $O(rN^2)$, where $N = N(M)$ is the total number of linear constraints given by (17). While this is substantially faster than straightforward Gaussian elimination, the problem is that the number of equations $N$ is often too large to make an $O(rN^2)$ computation feasible in practice. The following example sheds some light on the magnitude of this problem.

**Example 1a.** Let $\mathbb{C}_q(n, k)$ be a Reed-Solomon code of length $n = 255$ and dimension $k = 239$ over $\mathbb{F}_{256}$. A typical interpolation problem arising in algebraic soft-decision decoding [12] of $\mathbb{C}_q(n, k)$ might involve the following multiplicities:

| multiplicity    | 7    | 6   | 5   | 4  | 3  | 2  | 1  |
|----------------:|------|-----|-----|----|----|----|----|
| # of points     | 229  | 12  | 10  | 4  | 3  | 10 | 10 |
| # of constraints| 6412 | 252 | 150 | 40 | 18 | 30 | 10 |

(22)

for a total of 6192 linear equations. Using (18), we find that the corresponding value of $\delta^*$ is 1598. The $Y$-degree of a solution $\mathcal{Q}(X, Y)$ to this interpolation problem is $r = 6$, in view of (19).

Estimating the complexity of Koetter's interpolation algorithm as $rN^2$, we arrive at roughly $230 \times 10^6$ operations. In fact, the algorithm requires exactly $159.56 \times 10^6$ finite-field multiplications. This figure is precise: it was obtained by actually implementing the algorithm, and counting the number of finite-field multiplications in software. The fastest algorithm we found for the particular interpolation problem in (22) is due to Lee and O'Sullivan [13]. This algorithm takes exactly $45.37 \times 10^6$ finite field multiplications, which is still prohibitive. $\square$

This example illustrates a major problem with interpolation-based decoding. While, for a fixed maximal multiplicity, the decoding complexity is bounded by a polynomial in the length of the code, the actual complexity of solving the interpolation problem is prohibitively large in practice. In the next section, we present an algorithmic transformation that drastically reduces this complexity. Before describing this transformation in detail, let us use the following example to illustrate the savings in complexity that can be achieved with this method.

**Example 1b.** Consider again the interpolation problem in (22). Judiciously choosing the *re-encoding point set*, we can eliminate the $k = 239$ points with the highest multiplicities: the 229 points of multiplicity 7 as well as 10 of the 12 points of multiplicity 6. This leaves only 290 linear equations to solve, rather than the original 6912. This is a much more feasible task. In fact, using Koetter's algorithm to solve the reduced interpolation problem requires only $350 \times 10^3$ finite-field multiplications. The reduction in complexity, by a factor of 456, is augmented by a corresponding reduction in memory requirements, due to the fact that the polynomials operated upon during the interpolation procedure have much small degree. Specifically, the largest $X$-degree we need is about 40, instead of 1598. $\square$

## IV. A Complexity Reducing Transformation

Rather than seeking an efficient way to solve $\mathbf{IP}_{1,k-1}(\mathcal{P}, M)$, we will modify the interpolation problem itself, by means of a shift and a coordinate transformation. Our approach is similar to the re-encoding idea of Berlekamp and Welch [23].

### A. Re-Encoding and Shift

Given the point set $\mathcal{P} = \{(x_1, y_1), (x_2, y_2), \ldots, (x_s, y_s)\}$, we first identify some $k$ points $(x_{i_1}, y_{i_1}), (x_{i_2}, y_{i_2}), \ldots, (x_{i_k}, y_{i_k})$ in $\mathcal{P}$ such that $x_{i_1}, x_{i_2}, \ldots, x_{i_k} \in \mathcal{D}$ are all distinct. Define
$$\mathcal{R} \stackrel{\text{def}}{=} \big\{(x_{i_1}, y_{i_1}), (x_{i_2}, y_{i_2}), \ldots, (x_{i_k}, y_{i_k})\big\} \quad (23)$$

Observe that if the set $\mathcal{P}$ contains $n - e < k$ points with distinct $X$-coordinates, then the resulting interpolation polynomial $\mathcal{Q}(X, Y)$ will have at least $q^{e-(n-k)}$ factors of type $Y - f(X)$. This situation corresponds to $e > n - k$ erasures, in which case the transmitted codeword cannot be uniquely determined. This shows that, unless the interpolation problem $\mathbf{IP}_{1,k-1}(\mathcal{P}, M)$ is ill-conditioned by too many erasures, a **re-encoding point set** $\mathcal{R}$ with the required property always exists.

Note that there will usually be exponentially many ways to choose $\mathcal{R}$ from $\mathcal{P}$. As far as the theory developed in this paper is concerned, the choice of $\mathcal{R}$ is arbitrary. In practice, the set $\mathcal{R}$ will be chosen to consist of the points with the highest possible multiplicities (cf. Example 1). To simplify notation in what follows, we assume without loss of generality that $\mathcal{R}$ consists of the first $k$ points of $\mathcal{P}$, that is

$$\mathcal{R} = \{(x_1, y_1), (x_2, y_2), \ldots, (x_k, y_k)\}$$

The re-encoding point set $\mathcal{R}$ determines the unique ***re-encoding polynomial*** $e(X)$ of degree $< k$, defined by

$$e(x_i) = y_i \quad \text{for all } (x_i, y_i) \in \mathcal{R} \quad (24)$$

Observe that the codeword $\underline{c}'$ obtained by evaluating $e(X)$ at $\xi_1, \xi_2, \ldots, \xi_n$ agrees with the "given" values $y_1, y_2, \ldots, y_k$ at the $k$ positions corresponding to $x_1, x_2, \ldots, x_k$. Thus computing $e(X)$ is equivalent to re-encoding through $k$ given values at some $k$ positions. If these $k$ positions are consecutive and $\mathbb{C}_q(n,k)$ is cyclic, this can be achieved through division by the generator polynomial for $\mathbb{C}_q(n,k)$. Otherwise, such re-encoding is tantamount to correcting $n-k$ erasures in $\mathbb{C}_q(n,k)$. Various efficient algorithms for this purpose are known [15, 23].

Given the set $\mathcal{P} = \{(x_1, y_1), (x_2, y_2), \ldots, (x_s, y_s)\}$ and the re-encoding polynomial $e(X)$, we define

$$\mathcal{P}' \stackrel{\text{def}}{=} \{(x_1, y_1 - e(x_1)), (x_2, y_2 - e(x_2)), \ldots, (x_s, y_s - e(x_s))\}$$

Notice that, by the definition of $e(X)$ in (24), the first $k$ points in $\mathcal{P}'$ are of the form $(x_1, 0), (x_2, 0), \ldots, (x_k, 0)$.

**Lemma 2.** *Consider an arbitrary point $(\alpha, \beta) \in \mathcal{P}$ and the corresponding point $(\alpha, \beta') \in \mathcal{P}'$, where $\beta' = \beta - e(\alpha)$. Further, let $\mathcal{A}(X,Y)$ and $\mathcal{B}(X,Y)$ be arbitrary polynomials related by*

$$\mathcal{B}(X,Y) = \mathcal{A}(X, Y + e(X)) \quad (25)$$
$$\mathcal{A}(X,Y) = \mathcal{B}(X, Y - e(X)) \quad (26)$$

*Then*

$$\mu_{\alpha,\beta}(\mathcal{A}(X,Y)) = \mu_{\alpha,\beta'}(\mathcal{B}(X,Y)) \quad (27)$$

*Proof.* The proof is very similar to the proof of Theorem 1. Expand $\mathcal{B}(X,Y)$ in the basis $(X-\alpha)^i(Y-\beta')^j$ to write

$$\mathcal{B}(X,Y) = \sum_{i=0}^{\infty}\sum_{j=0}^{\infty} b_{i,j} (X-\alpha)^i (Y-\beta')^j \quad (28)$$

so that $\mu_{\alpha,\beta'}(\mathcal{B}(X,Y)) = \min_{b_{i,j} \neq 0}\{i+j\}$. Then use (26), (28) and the fact that $e(X) = (X-\alpha)h(X) + e(\alpha)$ for some polynomial $h(X)$, to express $\mathcal{A}(X+\alpha, Y+\beta)$ as follows:

$$\mathcal{A}(X+\alpha, Y+\beta) = \mathcal{B}(X+\alpha, Y+\beta - e(X+\alpha))$$
$$= \sum_{i=0}^{\infty}\sum_{j=0}^{\infty} b_{i,j} X^i \big(Y + e(\alpha) - e(X+\alpha)\big)^j$$
$$= \sum_{i=0}^{\infty}\sum_{j=0}^{\infty} b_{i,j} X^i \big(Y - Xh(X+\alpha)\big)^j \quad (29)$$

Since $\mu_{0,0}(X) = \mu_{0,0}(Y - Xh(X+\alpha)) = 1$, it follows immediately from (3), (4), (5), and (29) that

$$\mu_{\alpha,\beta}(\mathcal{A}(X,Y)) \geqslant \min_{b_{i,j}\neq 0}\{i+j\} = \mu_{\alpha,\beta'}(\mathcal{B}(X,Y))$$

The proof of the inequality $\mu_{\alpha,\beta'}(\mathcal{B}(X,Y)) \geqslant \mu_{\alpha,\beta}(\mathcal{A}(X,Y))$ is essentially identical to the above, and is omitted. ∎

**Remark.** It is not a coincidence that the proofs of Theorem 1 and Lemma 2 are so similar. Both results can be established in the general framework of birational maps between algebraic varieties [19]. However, there is no real need to invoke the heavy machinery of birational isomorphisms in this paper.

**Theorem 3.** *A polynomial $\mathcal{Q}(X,Y)$ is a solution to the interpolation problem $\mathbf{IP}_{1,k-1}(\mathcal{P}, M)$ if and only if the polynomial $\mathcal{Q}'(X,Y) = \mathcal{Q}(X, Y + e(X))$ is a solution to $\mathbf{IP}_{1,k-1}(\mathcal{P}', M)$.*

*Proof.* Let $\mathcal{I}(\mathcal{P}, M)$ denote the ideal of $\mathbb{F}_q[X,Y]$ consisting of all polynomials $\mathcal{A}(X,Y)$ that satisfy the interpolation constraints determined by $\mathcal{P}$ and $M$, namely such that

$$\mu_{x,y}(\mathcal{A}(X,Y)) \geqslant m_{x,y} \quad \text{for all } (x,y) \in \mathcal{P} \quad (30)$$

It follows from Lemma 2 that $\mathcal{Q}(X,Y) \in \mathcal{I}(\mathcal{P}, M)$ if and only if $\mathcal{Q}'(X,Y) = \mathcal{Q}(X, Y+e(X))$ is in $\mathcal{I}(\mathcal{P}', M)$. It remains to prove that $\mathcal{Q}(X,Y)$ is of minimal $(1,k-1)$-weighted degree in its ideal if and only if so is $\mathcal{Q}'(X,Y)$. Observe that

$$\deg_{1,k-1}\mathcal{Q}'(X,Y) = \deg_{1,k-1}\mathcal{Q}(X,Y)$$

This follows from the fact that $\deg e(X) \leqslant k-1$, and therefore $\deg_{1,k-1}Y = \deg_{1,k-1}(Y + e(X)) = k-1$. In general, it is easy to see that the transformations in (25) and (26) preserve the $(1,k-1)$-weighted degree. Now, assume to the contrary that $\mathcal{Q}'(X,Y)$ is minimal in $\mathcal{I}(\mathcal{P}', M)$, but $\mathcal{I}(\mathcal{P}, M)$ contains a polynomial $\mathcal{A}(X,Y)$ such that

$$\deg_{1,k-1}\mathcal{A}(X,Y) < \deg_{1,k-1}\mathcal{Q}(X,Y)$$

But then we can use (25) to produce a polynomial $\mathcal{B}(X,Y)$ in $\mathcal{I}(\mathcal{P}', M)$ whose $(1,k-1)$-weighted degree is strictly less than that of $\mathcal{Q}'(X,Y)$, a contradiction. ∎

It follows from Theorem 3 that instead of solving the interpolation problem $\mathbf{IP}_{1,k-1}(\mathcal{P}, M)$ directly, we may first compute a solution $\mathcal{Q}'(X,Y)$ to $\mathbf{IP}_{1,k-1}(\mathcal{P}', M)$ and then set

$$\mathcal{Q}(X,Y) := \mathcal{Q}'(X, Y - e(X))$$

Every solution $\mathcal{Q}(X,Y)$ to $\mathbf{IP}_{1,k-1}(\mathcal{P}, M)$ can be obtained in this manner. The next example illustrates the process in detail.

**Example 2a.** Let $\mathbb{F}_q = \mathbb{F}_8 = \{0, 1, \alpha, \alpha^2, \alpha^3, \alpha^4, \alpha^5, \alpha^6\}$, where $\alpha$ is a root of the primitive polynomial $X^3 + X + 1$. Take $k=2$ and $\mathcal{D} = \{1, \alpha, \alpha^2, \alpha^3\}$, and consider the Reed-Solomon code:

$$\mathbb{C}_8(4,2) = \{(f(1), f(\alpha), f(\alpha^2), f(\alpha^3)) : f(X) = a + bX\}$$

Suppose that the codeword $\underline{c} = (1, \alpha^4, \alpha^3, \alpha)$, corresponding to $f(X) = \alpha^6 + \alpha^2 X$, was transmitted. Further assume that, upon observing the channel output, the multiplicity assignment algorithm of [12] produces the following interpolation problem:

| $(x,y)$ | $(\alpha, \alpha^4)$ | $(\alpha^2, \alpha^6)$ | $(\alpha^2, \alpha^3)$ | $(\alpha^3, 1)$ | $(\alpha^3, \alpha)$ | $(1, \alpha)$ | $(1,1)$ |
|---|---|---|---|---|---|---|---|
| $m_{x,y}$ | 2 | 1 | 1 | 1 | 1 | 1 | 1 |

This roughly corresponds to receiving the hard-decision vector $\underline{v} = (\alpha, \alpha^4, \alpha^6, 1)$, with errors in positions 1, 2, 4. Since $k=2$, the $(1,k-1)$-weighted degree reduces to the total degree in this case. Hence, we have $\chi_{1,1}(\delta) = (\delta+1)(\delta+2)/2$. Since there are $N(M) = 9$ linear constraints, the corresponding value of $\delta^*$ is 3, and $r=3$ by (19). Koetter's interpolation algorithm thus initializes the Gröbner basis as $\mathscr{G}^{(0)} = \{1, Y, Y^2, Y^3\}$ and pro-ceeds through the 9 iterations detailed in Table I. Note that, at each iteration, the Gröbner basis polynomials are arranged in Table I in ascending order with respect to $\prec_k$. Thus the output of the algorithm is $\mathcal{Q}(X,Y) = \mathcal{G}_2^{(9)}(X,Y)$, given by

$$(1 + \alpha^5 X + \alpha X^3) + (\alpha^4 + X + X^2)Y + (\alpha^3 + X)Y^2 \quad (31)$$

It is easy to verify that $\mathcal{Q}(X,Y)$ satisfies all the interpolation constraints. Also, the polynomial $Y - (\alpha^6 + \alpha^2 X)$, which corresponds to the transmitted codeword, is a factor of $\mathcal{Q}(X,Y)$. In fact, the complete factorization of $\mathcal{Q}(X,Y)$ is as follows:

$$\mathcal{Q}(X,Y) = (\alpha^3 + X)(Y - (\alpha^6 + \alpha^2 X))(Y - (\alpha^5 + \alpha^6 X))$$

Now, let us apply the shift transformation of Theorem 3 with respect to the re-encoding point set

$$\mathcal{R} = \{(\alpha, \alpha^4), (\alpha^2, \alpha^6)\} \quad (32)$$

which consists of the first $k = 2$ points of $\mathcal{P}$. This set $\mathcal{R}$ determines the re-encoding polynomial $e(X) = \alpha^5 + \alpha^6 X$, with $e(\alpha) = \alpha^4$ and $e(\alpha^2) = \alpha^6$. The polynomial $e(X)$, in turn, determines the shifted interpolation point set $\mathcal{P}'$ given by

| $(x,y)$ | $(\alpha,0)$ | $(\alpha^2,0)$ | $(\alpha^2,\alpha^4)$ | $(\alpha^3,\alpha)$ | $(\alpha^3,1)$ | $(1,0)$ | $(1,\alpha^3)$ |
|---|---|---|---|---|---|---|---|
| $m_{x,y}$ | 2 | 1 | 1 | 1 | 1 | 1 | 1 |

The Koetter algorithm for the shifted problem $\mathbf{IP}_{1,k-1}(\mathcal{P}', M)$ proceeds as shown in in Table II, and produces the output

$$\mathcal{Q}'(X,Y) = (\alpha^4 + X + X^2)Y + (\alpha^3 + X)Y^2 \quad (33)$$

Again, it is easy to verify that $\mathcal{Q}'(X,Y)$ satisfies all the interpolation constraints of $\mathbf{IP}_{1,k-1}(\mathcal{P}', M)$, and factors as follows:

$$\mathcal{Q}'(X,Y) = (\alpha^3 + X)Y(Y - (\alpha + X))$$

We see that indeed, $\mathcal{Q}(X,Y) = \mathcal{Q}'(X, Y - (\alpha^5 + \alpha^6 X))$ as expected. Note that we could also obtain the $Y$-roots of $\mathcal{Q}(X,Y)$, namely $Y - (\alpha^6 + \alpha^2 X)$ and $Y - (\alpha^5 + \alpha^6 X)$, by first factoring $\mathcal{Q}'(X,Y)$ and then shifting its $Y$-roots. $\square$

### B. Coordinate Transformation

We are now ready to proceed with the complexity reducing transformations. For $m \in \mathbb{Z}$, define $[m]^+ = \max\{m, 0\}$.

**Lemma 4.** *The polynomial $\mathcal{A}(X,Y) = \sum_{j=0}^\infty a_j(X) Y^j$ passes through a point $(\alpha, 0)$ with multiplicity $m$ if and only if the univariate polynomials $a_j(X)$ are all divisible by $(X - \alpha)^{[m-j]^+}$.*

*Proof.* Expand $a_j(X)$ in the basis functions $(X - \alpha)^j$, that is write $a_j(X)$ as $a_j(X) = \sum_{i=0}^\infty a_{i,j}(X - \alpha)^i$. Then the expansion (1) of $\mathcal{A}(X,Y)$ at the point $(\alpha, 0)$ is given by

$$\mathcal{A}(X,Y) = \sum_{i,j=0}^\infty a_{i,j}(X - \alpha)^i (Y - 0)^j = \sum_{i,j=0}^\infty a_{i,j}(X - \alpha)^i Y^j$$

Clearly, the polynomial $a_j(X)$ is divisible by $(X - \alpha)^{[m-j]^+}$ if and only if $a_{i,j} = 0$ for all nonnegative $i < m - j$. This is just a reformulation of the definition of multiplicity. ∎

For ease of notation, let us henceforth denote the multiplicities $m_{x_1,y_1}, m_{x_2,y_2}, \ldots, m_{x_k,y_k}$ of the points in the re-encoding set $\mathcal{R}$ as $\nu_1, \nu_2, \ldots, \nu_k$. Thus $\nu_i = m_{x_i,y_i}$ for all $i = 1, 2, \ldots, k$. The following corollary is immediate from Lemma 4.

**Corollary 5.** *The polynomial $\mathcal{A}(X,Y) = \sum_{j=0}^\infty a_j(X) Y^j$ passes through the $k$ points $(x_1, 0), (x_2, 0), \ldots, (x_k, 0)$ with multiplicities $\nu_1, \nu_2, \ldots, \nu_k$ if and only if all the polynomials $a_j(X)$ are divisible by $\prod_{i=1}^k (X - x_i)^{[\nu_i - j]^+}$.*

From Corollary 5, we conclude that a solution $\mathcal{Q}'(X,Y)$ to the interpolation problem $\mathbf{IP}_{1,k-1}(\mathcal{P}', M)$ must have the form

$$\mathcal{Q}'(X,Y) = \sum_{j=0}^r \left( q_j(X) \prod_{i=1}^k (X - x_i)^{[\nu_i - j]^+} \right) Y^j \quad (34)$$

for some polynomials $q_0(X), q_1(X), \ldots, q_r(X)$. This observation makes it possible to reduce the number of iterations in the interpolation algorithm by *pre-solving* for the first $k$ interpolation points $(x_1, 0), (x_2, 0), \ldots, (x_k, 0)$. Specifically, in lieu of (21), we initialize the Gröbner-basis polynomials as follows:

$$\mathcal{G}_j^{(0)}(X,Y) = Y^j \prod_{i=1}^k (X - x_i)^{[\nu_i - j]^+} \text{ for } j = 0, 1, \ldots, r \quad (35)$$

By Corollary 5, these polynomials *already satisfy* all the interpolation constraints determined by $(x_1, 0), (x_2, 0), \ldots, (x_k, 0)$. Thus it remains to enforce only those interpolation constraints that correspond to the other points in $\mathcal{P}'$.

**Example 2b.** Consider again the re-encoding transformation in Example 2a. Recall that $k = 2$, and the relevant interpolation points are $(x_1, 0) = (\alpha, 0)$ and $(x_2, 0) = (\alpha^2, 0)$, with multiplicities $\nu_1 = 2$ and $\nu_2 = 1$ respectively. According to (35), let us initialize the Gröbner basis as follows:

$$\begin{aligned}\mathcal{G}_0^{(0)}(X,Y) &:= (X - \alpha)^2(X - \alpha^2) \\ \mathcal{G}_1^{(0)}(X,Y) &:= Y(X - \alpha) \\ \mathcal{G}_2^{(0)}(X,Y) &:= Y^2 \\ \mathcal{G}_3^{(0)}(X,Y) &:= Y^3\end{aligned} \quad (36)$$

Comparing with Table II, we see that these are precisely the polynomials produced by the Koetter algorithm after the first 4 iterations, starting with the standard initialization $\{1, Y, Y^2, Y^3\}$. Thus, given the initialization in (36), it remains to proceed with the last 5 iterations of the algorithm, exactly as before. $\square$

This is nice but not enough, since most of the computation takes place in the later iterations of the interpolation algorithm (cf. Tables I and II). To further reduce the interpolation complexity, let us introduce the auxiliary polynomials $g(X), \psi(X)$, and the "tail" polynomials $t_j(X)$, defined as follows:

$$g(X) \stackrel{\text{def}}{=} \prod_{i=1}^k (X - x_i) \quad (37)$$

$$\psi(X) \stackrel{\text{def}}{=} \prod_{i=1}^k (X - x_i)^{\nu_i} \quad (38)$$

$$t_j(X) \stackrel{\text{def}}{=} \prod_{i=1}^k (X - x_i)^{[j - \nu_i]^+} \text{ for } j = 0, 1, \ldots, r \quad (39)$$

Combining these definitions with (34), we can express a solution $\mathcal{Q}'(X,Y)$ to the interpolation problem $\mathbf{IP}_{1,k-1}(\mathcal{P}', M)$ as

$$\mathcal{Q}'(X,Y) = \psi(X) \sum_{j=0}^r q_j(X) t_j(X) \left( \frac{Y}{g(X)} \right)^j \quad (40)$$

Computing $\mathcal{Q}'(X,Y)$ and thereby solving both $\mathbf{IP}_{1,k-1}(\mathcal{P}', M)$ and $\mathbf{IP}_{1,k-1}(\mathcal{P}, M)$ (in view of Theorem 3) thus reduces to find-ing $q_0(X), q_1(X), \ldots, q_r(X)$ in (40). The following two propositions show that computing $q_0(X), q_1(X), \ldots, q_r(X)$ is tantamount to solving a much smaller interpolation problem!

**Proposition 6.** *Let $(\alpha, \beta) \in \mathcal{P}'$ be such that $g(\alpha) \neq 0$. Then the polynomial $\mathcal{Q}'(X, Y)$, as defined in (40), passes through $(\alpha, \beta)$ with multiplicity $m$ if and only if the polynomial*

$$\mathcal{H}(X, Y) \stackrel{\text{def}}{=} \sum_{j=0}^{r} q_j(X) \, t_j(X) \, Y^j \qquad (41)$$

*passes with multiplicity $m$ through the point $(\alpha, \gamma) = \varphi_g(\alpha, \beta)$, where $\varphi_g$ is the birational mapping defined in (7).*

*Proof.* In order to prove the proposition, it is necessary and sufficient to show that

$$\mu_{\alpha,\beta}(\mathcal{Q}'(X, Y)) = \mu_{\alpha,\gamma}(\mathcal{H}(X, Y))$$

Note that in view of (37) and (38), we have $\psi(\alpha) \neq 0$ whenever $g(\alpha) \neq 0$. Hence $\mu_{\alpha,\beta}(\psi(X)) = 0$. Let us define

$$\mathcal{U}(X, Y) \stackrel{\text{def}}{=} \frac{\mathcal{Q}'(X, Y)}{\psi(X)} = \sum_{j=0}^{r} q_j(X) \, t_j(X) \left(\frac{Y}{g(X)}\right)^j \qquad (42)$$

Then $\mu_{\alpha,\beta}(\mathcal{Q}'(X, Y)) = \mu_{\alpha,\beta}(\mathcal{U}(X, Y))$ in view of (6). Therefore, it is enough to show that

$$\mu_{\alpha,\beta}(\mathcal{U}(X, Y)) = \mu_{\alpha,\gamma}(\mathcal{H}(X, Y))$$

But this follows from Theorem 1, since $\mathcal{U}(X, Y)$ does not have a pole at $(\alpha, \beta)$ and $\mathcal{H}(X, Y) = \Phi_g(\mathcal{U}(X, Y))$ in view of (41) and (42), where $\Phi_g$ is the map introduced in Definition 4. ∎

Proposition 6 will be useful for those points $(\alpha, \beta) \in \mathcal{P}'$ for which $g(\alpha) \neq 0$. However, $\mathcal{P}'$ may also contain points whose $X$-coordinate coincides with the $X$-coordinate $x_i$ of one of the points in the re-encoding set $\mathcal{R}$. For such points $(\alpha, \beta) \in \mathcal{P}'$, we have $g(\alpha) = 0$ by (37). In this case, we will use Proposition 7.

**Proposition 7.** *Let the point $(\alpha, \beta) \in \mathcal{P}'$ be such that $\alpha = x_i$ for some point $(x_i, y_i)$ in the re-encoding point set $\mathcal{R}$. Then the polynomial $\mathcal{Q}'(X, Y)$, as defined in (40), passes through $(\alpha, \beta)$ with multiplicity $m$ if and only if the polynomial*

$$\mathcal{H}'(X, Y) \stackrel{\text{def}}{=} \sum_{j=0}^{r} q_j(X) \, (X-\alpha)^{\nu_i - j} \, t_j(X) \, Y^j \qquad (43)$$

*passes with multiplicity $m$ through the point $(\alpha, \gamma)$, with $\gamma$ given by $\gamma = \beta/g'(\alpha)$, where $g'(X)$ is the derivative of $g(X)$ in (37).*

*Proof.* Again, to prove the proposition, it is necessary and sufficient to show that

$$\mu_{\alpha,\beta}(\mathcal{Q}'(X, Y)) = \mu_{\alpha,\gamma}(\mathcal{H}'(X, Y))$$

In order to do so, let us first introduce the auxiliary polynomials $h(X)$ and $\phi(X)$ defined as follows:

$$h(X) \stackrel{\text{def}}{=} \frac{g(X)}{X - \alpha} = \prod_{\substack{i=1 \\ x_i \neq \alpha}}^{k} (X - x_i) \qquad (44)$$

$$\phi(X) \stackrel{\text{def}}{=} \frac{\psi(X)}{(X - \alpha)^{\nu_i}} = \prod_{\substack{i=1 \\ x_i \neq \alpha}}^{k} (X - x_i)^{\nu_i} \qquad (45)$$

Observe that, since $\alpha$ is a simple root of $g(X)$ by (37), we have $h(\alpha) = g'(\alpha)$, where $g'(X)$ is the (first-order Hasse) derivative of $g(X)$. Hence $\gamma$ can be written as $\beta/h(\alpha)$ and we have $(\alpha, \gamma) = \varphi_h(\alpha, \beta)$, where $\varphi_h$ is the birational mapping in (7) defined with respect to $h(X)$ rather than $g(X)$. Also observe that $\phi(\alpha) \neq 0$ in view of (45). Now, let us rewrite the expression (40) for $\mathcal{Q}'(X, Y)$ as follows:

$$\mathcal{Q}'(X, Y) = \phi(X) \sum_{j=0}^{r} q_j(X) \, (X-\alpha)^{\nu_i - j} \, t_j(X) \left(\frac{Y}{h(X)}\right)^j$$

The rest of the proof is exactly the same as that of Proposition 6. We define the rational function

$$\mathcal{V}(X, Y) \stackrel{\text{def}}{=} \frac{\mathcal{Q}'(X, Y)}{\phi(X)}$$

and observe that $\mu_{\alpha,\beta}(\mathcal{Q}'(X, Y)) = \mu_{\alpha,\beta}(\mathcal{V}(X, Y))$ in view of (6) and the fact that $\mu_{\alpha,\beta}(\phi(X)) = 0$. Further, $\mathcal{V}(X, Y)$ doesn't have a pole at $(\alpha, \beta)$ and $\mathcal{H}'(X, Y) = \Phi_h(\mathcal{V}(X, Y))$, where $\Phi_h$ is the map in (9) defined with respect to $h(X)$. The proposition now follows immediately from Theorem 1. ∎

Observe that the polynomials $\mathcal{H}(X, Y)$ and $\mathcal{H}'(X, Y)$ defined in (41) and (43) are related via the transformation

$$\mathcal{H}'(X, Y) = (X-\alpha)^{\nu_i} \mathcal{H}\left(X, \frac{Y}{X-\alpha}\right)$$

Therefore $\mathcal{H}'(X, Y)$ passes through a given point $(\alpha, \gamma)$ with multiplicity $m$ if and only if

$$\mu_{\alpha,\gamma}\left((X-\alpha)^{\nu_i} \mathcal{H}\left(X, \frac{Y}{X-\alpha}\right)\right) \geq m \qquad (46)$$

It follows from Corollary 5, Proposition 6, Proposition 7, and (46) that all the interpolation constraints of the shifted interpolation problem $\mathbf{IP}_{1,k-1}(\mathcal{P}', M)$ can be expressed in terms of constraints on the polynomial $\mathcal{H}(X, Y)$ in (41).

It will be convenient to recast these constraints on $\mathcal{H}(X, Y)$ in terms of a point set $\mathcal{P}^*$ and a corresponding set of multiplicities $M^*$. Given the original point set $\mathcal{P}$ in (15) and the re-encoding point set $\mathcal{R}$ in (23), let us define

$\mathcal{S} \stackrel{\text{def}}{=}$ the set of points in $\mathcal{P} \setminus \mathcal{R}$ whose $X$-coordinates differ from those of the re-encoding points

$\mathcal{T} \stackrel{\text{def}}{=}$ the set of points in $\mathcal{P} \setminus \mathcal{R}$ whose $X$-coordinates coincide with those of the re-encoding points

Thus the sets $\mathcal{S}$ and $\mathcal{T}$ form a partition of $\mathcal{P} \setminus \mathcal{R}$. The coordinate transformation consists of converting $\mathcal{P} \setminus \mathcal{R}$ into the ***reduced point set*** $\mathcal{P}^* = \{(x_i, z_i) : (x_i, y_i) \in \mathcal{P} \setminus \mathcal{R}\}$, where

$$z_i \stackrel{\text{def}}{=} \begin{cases} \dfrac{y_i - e(x_i)}{g(x_i)} & \text{if } (x_i, y_i) \in \mathcal{S} \\ \dfrac{y_i - e(x_i)}{g'(x_i)} & \text{if } (x_i, y_i) \in \mathcal{T} \end{cases} \qquad (47)$$

We let $\mathcal{S}^*$ and $\mathcal{T}^*$ denote the sets of points in $\mathcal{P}^*$ transformed from the points in $\mathcal{S}$ and $\mathcal{T}$, respectively. Further, let $M^* \subset M$ denote the multiplicities of the points in $\mathcal{P}^*$, which are equal to the multiplicities of the original points in $\mathcal{P} \setminus \mathcal{R}$. Thus

$$M^* = \{m_{x_{k+1}, y_{k+1}}, m_{x_{k+2}, y_{k+2}}, \ldots, m_{x_s, y_s}\}$$

Propositions 6 and 7, and the coordinate transformation (47), are key to converting the original interpolation problem into the ***reduced interpolation problem***, defined in the next subsection.

## C. Reduced Interpolation Problem

Before introducing the reduced interpolation problem in Definition 6, we need to establish one more result.

**Lemma 8.** *Let $\mathcal{Q}'(X,Y)$ be a solution to the interpolation problem $\mathbf{IP}_{1,k-1}(\mathcal{P}', M)$. Let $\psi(X)$ and $\mathcal{H}(X,Y)$ be the polynomials defined in (38) and (41), respectively. Then*

$$\deg_{1,k-1} \mathcal{Q}'(X,Y) = \deg \psi(X) + \deg_{1,-1} \mathcal{H}(X,Y)$$

*Proof.* Reformulating (40) once again, let us write $\mathcal{Q}'(X,Y)$ as $\mathcal{Q}'(X,Y) = \sum_{j=0}^{r} p_j(X) Y^j$, where

$$p_j(X) \stackrel{\text{def}}{=} \frac{\psi(X) q_j(X) t_j(X)}{g(X)^j} \quad \text{for } j = 0, 1, \ldots, r$$

Note that $p_j(X)$ is a polynomial (rather than a rational function) for all $j$, and its degree is given by

$$\deg p_j(X) = \deg \psi(X) + \deg q_j(X) t_j(X) - jk$$

since $\deg g(X) = k$ by (37). Since the $r$ terms in $\sum_{j=0}^{r} p_j(X) Y^j$ have distinct $Y$-degrees, it follows that

$$\begin{aligned}
\deg_{1,k-1} \mathcal{Q}'(X,Y) &= \\
&= \max_{0 \leqslant j \leqslant r} \left\{ \deg p_j(X) + j(k-1) \right\} \\
&= \deg \psi(X) + \max_{0 \leqslant j \leqslant r} \left\{ \deg q_j(X) t_j(X) - j \right\} \\
&= \deg \psi(X) + \deg_{1,-1} \mathcal{H}(X,Y) \quad \blacksquare
\end{aligned}$$

Lemma 8 is the reason we use the $(1,-1)$-weighted degree, rather than $(1, k-1)$-weighted degree, in the definition below.

**Definition 6 (Reduced interpolation problem).** *Given the set of points $\mathcal{P}^* = \{(x_1, z_1), (x_2, z_2), \ldots, (x_{s-k}, z_{s-k})\}$, partitioned into subsets $\mathcal{S}^*$ and $\mathcal{T}^*$, and the corresponding set of multiplicities $M^* = \{m_{x_1,z_1}, m_{x_2,z_2}, \ldots, m_{x_{s-k},z_{s-k}}\}$, the reduced interpolation problem consists of computing a nonzero polynomial $\mathcal{H}(X,Y)$ which can be expressed in the form*

$$\mathcal{H}(X,Y) = \sum_{j=0}^{\infty} q_j(X) t_j(X) Y^j \quad (48)$$

*and satisfies*

$$\mu_{x_i, z_i}(\mathcal{H}(X,Y)) \geqslant m_{x_i, z_i} \quad \forall (x_i, z_i) \in \mathcal{S}^* \quad (49)$$

$$\mu_{x_i, z_i}\left( (X - x_i)^{\nu_i} \mathcal{H}\left(X, \frac{Y}{X - x_i}\right) \right) \geqslant m_{x_i, z_i} \quad \forall (x_i, z_i) \in \mathcal{T}^* \quad (50)$$

*such that $\deg_{1,-1} \mathcal{H}(X,Y)$ is minimal among all bivariate polynomials that satisfy the constraints (48), (49), and (50).*

We shall refer to the reduced interpolation problem above as $\mathbf{RIP}_{1,-1}(\mathcal{P}^*, M^*)$. The following theorem summarizes our results and establishes the connection between $\mathbf{RIP}_{1,-1}(\mathcal{P}^*, M^*)$ and the original interpolation problem $\mathbf{IP}_{1,k-1}(\mathcal{P}, M)$.

**Theorem 9.** *Let $\mathcal{H}(X,Y)$ be a solution to the reduced interpolation problem $\mathbf{RIP}_{1,-1}(\mathcal{P}^*, M^*)$. Then a solution to the original interpolation problem $\mathbf{IP}_{1,k-1}(\mathcal{P}, M)$ is given by*

$$\mathcal{Q}(X,Y) = \psi(X)\, \mathcal{H}\left(X, \frac{Y - e(X)}{g(X)}\right) \quad (51)$$

*Proof.* Let $\mathcal{Q}'(X,Y) \stackrel{\text{def}}{=} \psi(X) \mathcal{H}(X, Y/g(X))$. It is enough to show that $\mathcal{Q}'(X,Y)$ is a solution to the shifted interpolation problem $\mathbf{IP}_{1,k-1}(\mathcal{P}', M)$, since then the claim follows by Theorem 3. Proposition 6 and Proposition 7 along with (49) and (50) imply that $\mathcal{Q}'(X,Y)$ passes through the points $(x_{k+1}, y_{k+1})$, $(x_{k+2}, y_{k+2}), \ldots, (x_s, y_s)$ in $\mathcal{P}'$ with the prescribed multiplicities. Using the definitions of $g(X)$, $\psi(X)$, and $t_j(X)$ in (37), (38), and (39), we can express $\mathcal{Q}'(X,Y)$ as follows:

$$\mathcal{Q}'(X,Y) = \psi(X) \mathcal{H}\left(X, \frac{Y}{g(X)}\right) \quad (52)$$

$$= \sum_{j=0}^{r} q_j(X) \psi(X) t_j(X) \left(\frac{Y}{g(X)}\right)^j \quad (53)$$

$$= \sum_{j=0}^{r} q_j(X) \prod_{i=1}^{k} (X - x_i)^{[\nu_i - j]^+} Y^j \quad (54)$$

where the last equality follows from the fact that, for all integers $a$ and $b$, we have $a - b = [a - b]^+ - [b - a]^+$. It follows from (54) and Corollary 5 that $\mathcal{Q}'(X,Y)$ also passes through the points $(x_1, 0), (x_2, 0), \ldots, (x_k, 0)$ in $\mathcal{P}'$ with multiplicities $\nu_1, \nu_2, \ldots, \nu_k$. Thus $\mathcal{Q}'(X,Y)$ is in the ideal $\mathcal{I}(\mathcal{P}', M)$ of polynomials that satisfy all the constraints of $\mathbf{IP}_{1,k-1}(\mathcal{P}', M)$.

It remains to prove that $\mathcal{Q}'(X,Y)$ is of minimal $(1, k-1)$-weighted degree in this ideal. Assume to the contrary that there exists a polynomial $\mathcal{A}(X,Y) \in \mathcal{I}(\mathcal{P}', M)$ such that

$$\deg_{1,k-1} \mathcal{A}(X,Y) < \deg_{1,k-1} \mathcal{Q}'(X,Y) \quad (55)$$

Combining Corollary 5 with the definitions of $g(X)$, $\psi(X)$, and $t_j(X)$, we conclude that $\mathcal{A}(X,Y)$ can be expressed as

$$\mathcal{A}(X,Y) = \psi(X) \sum_{j=0}^{\infty} a_j(X) t_j(X) \left(\frac{Y}{g(X)}\right)^j \quad (56)$$

Given (56), we can construct a solution to $\mathbf{RIP}_{1,-1}(\mathcal{P}^*, M^*)$ as follows. Define $\mathcal{B}(X,Y) = \mathcal{A}(X, Yg(X))/\psi(X)$, namely

$$\mathcal{B}(X,Y) = \sum_{j=0}^{\infty} a_j(X) t_j(X) Y^j \quad (57)$$

Note that (57) is of the form (48). Furthermore, Propositions 6 and 7 imply that $\mathcal{B}(X,Y)$ satisfies (49) and (50). By Lemma 8

$$\deg_{1,k-1} \mathcal{A}(X,Y) = \deg \psi(X) + \deg_{1,-1} \mathcal{B}(X,Y)$$

which, in conjunction with (55), implies that $\deg_{1,-1} \mathcal{B}(X,Y)$ is strictly smaller than $\deg_{1,-1} \mathcal{H}(X,Y)$, a contradiction. $\blacksquare$

The reduced interpolation problem $\mathbf{RIP}_{1,-1}(\mathcal{P}^*, M^*)$ can be solved using a variant of Koetter's interpolation algorithm, described in the previous section. We need to make the following three modifications. First, in view of (48), we need to initialize the Gröbner-basis polynomials as follows:

$$\mathcal{G}_j^{(0)}(X,Y) := t_j(X) Y^j = Y^j \prod_{i=1}^{k} (X - x_i)^{[j - \nu_i]^+} \quad (58)$$

for $j = 0, 1, \ldots, r$. This initialization guarantees that the Gröbner-basis polynomials satisfy (48) through *all the iterations* of the algorithm. Second, in view of (50), in those iterations that deal with points $(\alpha, \gamma) \in \mathcal{T}^*$, the discrepancies $\Delta_j$ need to be defined differently. Specifically, suppose that $\alpha = x_i$ for some point $(x_i, y_i)$ in the re-encoding point set $\mathcal{R}$, and let $\nu_i$ denote the multiplicity of this point (as before). Then the discrepancies should be computed as follows:

$$\Delta_j := \mathbf{coef}\left(X^{\nu_i}\mathcal{G}_j\left(X+\alpha, \frac{Y+\gamma}{X}\right); X^a Y^b\right) \quad (59)$$

for $j = 0, 1, \ldots, r$. Third, instead of the monomial order $\prec_k$ defined in (20), we need to order the bivariate monomials as follows: we say that $X^a Y^b \prec_{-1} X^i Y^j$ iff

$$\bigl(a - b < i - j\bigr) \quad \text{or} \quad \bigl(a - b = i - j \text{ and } b < j\bigr) \quad (60)$$

Note that $\prec_{-1}$ is *not* a monomial order, since it does not have the well-ordering property (there is no smallest monomial under $\prec_{-1}$). This, however, has no effect on Koetter's algorithm, and we may safely replace $\prec_k$ with $\prec_{-1}$ throughout.

**Example 2c.** Consider again the re-encoding transformation in Example 2a. Recall that $k = 2$, and the re-encoding set $\mathcal{R}$ consists of $(x_1, y_1) = (\alpha, \alpha^4)$ and $(x_2, y_2) = (\alpha^2, \alpha^6)$, with multiplicities $\nu_1 = 2$ and $\nu_2 = 1$ respectively. Thus

$$g(X) := \prod_{i=1}^{2}(X - x_i) = (X - \alpha)(X - \alpha^2) \quad (61)$$

$$\psi(X) := \prod_{i=1}^{2}(X - x_i)^{\nu_i} = (X - \alpha)^2(X - \alpha^2) \quad (62)$$

according to (37) and (38), respectively. Since $r = 3$, there are four tail polynomials $t_0(X), t_1(X), t_2(X), t_3(X)$. However, the first two are trivial: $t_0(X) = t_1(X) = 1$. Indeed, let us define $\nu_{\min} = \min_{1 \leq i \leq k} \nu_i$. Then the definition of tail polynomials in (39) implies that $t_j(X) = 1$ for $j \leq \nu_{\min}$. In fact, if $\nu_{\min} \geq r$, then *all* the tail polynomials are trivial, and the constraint (48) on the form of a solution to $\mathbf{RIP}_{1,-1}(\mathcal{P}^*, M^*)$ becomes vacuous. In this example, however, we have $\nu_{\min} = 1$, and

$$t_2(X) = X - \alpha^2$$
$$t_3(X) = (X - \alpha)(X - \alpha^2)^2 = \alpha^5 + \alpha^4 X + \alpha X^2 + X^3$$

Thus, according to (58), the Gröbner-basis polynomials should be initialized as follows

$$\mathcal{G}_0^{(0)}(X, Y) := t_0(X) = 1$$
$$\mathcal{G}_1^{(0)}(X, Y) := t_1(X)Y = Y$$
$$\mathcal{G}_2^{(0)}(X, Y) := t_2(X)Y^2 = (X + \alpha^2)Y^2$$
$$\mathcal{G}_3^{(0)}(X, Y) := t_2(X)Y^3 = (\alpha^5 + \alpha^4 X + \alpha X^2 + X^3)Y^3$$

Next, let us compute the reduced point set $\mathcal{P}^*$ using the coordinate transformation defined in (47). The partition of the five points in $\mathcal{P} \setminus \mathcal{R}$ into the sets $\mathcal{S}$ and $\mathcal{T}$ is given by

$$\mathcal{S} = \left\{(\alpha^3, 1), (\alpha^3, \alpha), (1, \alpha), (1, 1)\right\}$$
$$\mathcal{T} = \left\{(\alpha^2, \alpha^3)\right\}$$

since $g(\alpha^2) = 0$, while $g(\alpha^3) = \alpha^5$ and $g(1) = \alpha^2$ are nonzero. Also $g'(\alpha^2) = \alpha^4$. Dividing the shifted interpolation points in Example 2b by $\alpha^5$, $\alpha^2$, or $\alpha^4$, as appropriate, we compute

$$\mathcal{S}^* = \left\{(\alpha^3, \alpha^2), (\alpha^3, \alpha^3), (1, 0), (1, \alpha)\right\}$$
$$\mathcal{T}^* = \left\{(\alpha^2, 1)\right\}$$

The multiplicity of all the points in $\mathcal{P}^* = \mathcal{S}^* \cup \mathcal{T}^*$ is 1, as in Example 2b. This determines all the interpolation constraints of $\mathbf{RIP}_{1,-1}(\mathcal{P}^*, M^*)$, and Koetter's interpolation algorithm, subject to the modifications (59) and (60), now proceeds through the 5 iterations detailed in Table III. Again, at each iteration, the Gröbner basis polynomials are arranged in Table III in ascending order with respect to $\prec_{-1}$. Thus the output of the algorithm is the polynomial $\mathcal{H}(X, Y) = \mathcal{G}_2^{(5)}(X, Y)$, given by

$$\mathcal{H}(X, Y) = (\alpha^3 + X)Y + (\alpha^5 + \alpha^5 X + X^2)Y^2 \quad (63)$$

It is easy to verify that $\psi(X)\mathcal{H}\bigl(X, Y/g(X)\bigr)$, where $g(X)$ and $\psi(X)$ are given by (61) and (62), is precisely the polynomial $\mathcal{Q}'(X, Y)$ in (33). It can be also verified directly that

$$\mathcal{Q}(X, Y) = (X-\alpha)^2(X-\alpha^2)\,\mathcal{H}\left(X, \frac{Y - \alpha^5 - \alpha^6 X}{(X-\alpha)(X-\alpha^2)}\right)$$

is precisely the solution to the original interpolation problem found in Example 2a (cf. (31) or the last row of Table I). □

**Remark.** Strictly speaking, the set $\mathcal{J}(\mathcal{P}^*, M^*)$ of all bivariate polynomials over $\mathbb{F}_q$ that satisfy the constraints (48), (49), (50) is no longer an ideal of $\mathbb{F}_q[X, Y]$. The problem is with the constraint (48). For example, if $\mathcal{A}(X, Y) \in \mathcal{J}(\mathcal{P}^*, M^*)$ then $Y\mathcal{A}(X, Y)$ is not necessarily in $\mathcal{J}(\mathcal{P}^*, M^*)$ since it may not be possible to express $Y\mathcal{A}(X, Y)$ in the form of (48). Nevertheless, if we take the intersection of $\mathcal{J}(\mathcal{P}^*, M^*)$ with the set of polynomials in $\mathbb{F}_q[X, Y]$ whose $Y$-degree is at most $r$, then this intersection is a module over $\mathbb{F}_q[X]$. What the (modified) Koetter algorithm computes is the Gröbner basis for this module.

In summary, we emphasize that although $\mathbf{RIP}_{1,-1}(\mathcal{P}^*, M^*)$ appears to be much more convoluted than the original problem $\mathbf{IP}_{1,k-1}(\mathcal{P}, M)$, its complexity is often orders of magnitude lower. This is due to the fact that we *do not even need to consider* the first $k$ points of $\mathcal{P}$ in computing $\mathcal{H}(X, Y)$. In other words, these $k$ interpolation points (which can be selected to have the largest multiplicities) effectively disappear. The complexity savings are not apparent in the toy example studied in this section. However, they become prominent for long high-rate Reed-Solomon codes of practical interest (cf. Example 1).

## V. THE FACTORIZATION PROCEDURE

The reduction in complexity achieved in the previous section would be much less significant if one had to actually use (51) to compute a solution to the original interpolation problem (or the shifted interpolation problem), and then factor this solution in order to recover the transmitted codeword. The problem is that (51) involves multiplication by $\psi(X)$, which is a large polynomial, usually *much larger* than the solution $\mathcal{H}(X, Y)$ to $\mathbf{RIP}_{1,-1}(\mathcal{P}^*, M^*)$ computed by the Koetter algorithm. For instance, in the situation of Example 1, typical for Reed-Solomon codes of practical interest, we have $\deg \psi(X) = 1663$ whereas $\deg_{1,0} \mathcal{H}(X, Y) = 44$ and $\deg_{0,1} \mathcal{H}(X, Y) = 6$. Thus in factoring $\mathcal{H}(X, Y)$, one deals with polynomials of degree at most 44, whereas factoring $\mathcal{Q}'(X, Y) = \psi(X)\mathcal{H}(X, Y/g(X))$ involves processing polynomials of degree 1598.

Fortunately, the following theorem shows that instead of factoring $\mathcal{Q}(X, Y)$ or $\mathcal{Q}'(X, Y)$, we can directly factor the much smaller polynomial $\mathcal{H}(X, Y)$ to find the transmitted codeword.

**Theorem 10.** *Let $\mathcal{H}(X,Y)$ be a solution to the reduced interpolation problem $\mathbf{RIP}_{1,-1}(\mathcal{P}^*, M^*)$ and let*

$$\mathcal{Q}(X,Y) = \psi(X)\,\mathcal{H}\!\left(X, \frac{Y-e(X)}{g(X)}\right) \quad (64)$$

*be the corresponding solution to the original interpolation problem $\mathbf{IP}_{1,k-1}(\mathcal{P}, M)$. If $\mathcal{Q}(X,Y)$ has a factor $Y-f(X)$, where $\deg f(X) < k$, then $\mathcal{H}(X,Y)$ has a factor of the form*

$$Y\sigma(X) - \omega(X) = \sigma(X)\!\left(Y - \frac{\omega(X)}{\sigma(X)}\right) \quad (65)$$

*where $\sigma(X)$ and $\omega(X)$ are the error-locator and error-evaluator polynomials for the re-encoding positions with respect to $f(X)$.*

*Proof.* Recall that the re-encoding point set $\mathcal{R}$ consists of the $k$ points $(x_1,y_1), (x_2,y_2), \ldots, (x_k,y_k)$ with $x_1, x_2, \ldots, x_k$ being distinct elements of $\mathcal{D}$. By the **error-locator polynomial** for the re-encoding positions with respect to $f(X)$, we mean a monic polynomial $\sigma(X)$ with $\sigma(x_i) = 0$ whenever $f(x_i) \neq y_i$ and $\sigma(x) \neq 0$ for all other $x \in \mathbb{F}_q$. In other words, let

$$\mathcal{E} \stackrel{\text{def}}{=} \bigl\{i \in \{1,2,\ldots,k\} \;:\; y_i \neq f(x_i)\bigr\} \quad (66)$$

be the set of error locations among the re-encoding positions. Then the error-locator polynomial $\sigma(X)$ is given by

$$\sigma(X) \stackrel{\text{def}}{=} \prod_{i \in \mathcal{E}} (X - x_i) \quad (67)$$

Further, let $\mathcal{E}^c$ be the complement of $\mathcal{E}$ in the set $\{1,2,\ldots,k\}$, and define the polynomial

$$\lambda(X) \stackrel{\text{def}}{=} \prod_{i \in \mathcal{E}^c}(X - x_i) \quad (68)$$

By an **error-evaluator polynomial** for the re-encoding positions with respect to $f(X)$, we mean a polynomial $\omega(X)$ such that whenever $f(x_i) \neq y_i$, their difference is given by

$$f(x_i) - y_i = \omega(x_i)\lambda(x_i) = \omega(x_i)\frac{g'(x_i)}{\sigma'(x_i)} \quad (69)$$

where $g'(X)$ and $\sigma'(X)$ are the (first-order Hasse) derivatives of the polynomials $g(X)$ in (37) and $\sigma(X)$ in (67).

Now suppose that $\mathcal{Q}(X,Y)$ has $Y - f(X)$ as a factor. That is, there is a polynomial $\mathcal{A}(X,Y)$ such that

$$\mathcal{Q}(X,Y) = \bigl(Y - f(X)\bigr)\mathcal{A}(X,Y) \quad (70)$$

Let $e(X)$ be the re-encoding polynomial defined in (24), and let $\mathcal{Q}'(X,Y) = \mathcal{Q}(X, Y+e(X))$ be the corresponding solution to the shifted interpolation problem. It follows from (70) that

$$\mathcal{Q}'(X,Y) = \bigl(Y - \eta(X)\bigr)\mathcal{A}(X,Y+e(X)) \quad (71)$$

where $\eta(X) \stackrel{\text{def}}{=} f(X) - e(X)$. If $\mathcal{H}(X,Y)$ is the corresponding solution to the reduced interpolation problem, then

$$\psi(X)\,\mathcal{H}\!\left(X, \frac{Y}{g(X)}\right) = \bigl(Y - \eta(X)\bigr)\mathcal{A}(X, Y+e(X)) \quad (72)$$

Finally, applying the birational transformation $\Phi_g$ in (9) to both sides of (72), we obtain

$$\psi(X)\,\mathcal{H}(X,Y) = \bigl(Yg(X) - \eta(X)\bigr)\mathcal{B}(X,Y) \quad (73)$$

where $\mathcal{B}(X,Y) = \mathcal{A}(X, Yg(X) + e(X))$. It follows that the polynomial $\psi(X)\mathcal{H}(X,Y)$ is divisible by $Yg(X) - \eta(X)$.

Now observe that $\eta(x_i) = f(x_i) - e(x_i) = f(x_i) - y_i = 0$ for all $i \in \mathcal{E}^c$. Therefore $\eta(X)$ is divisible by $\lambda(X)$, and we can express $\eta(X)$ as $\omega(X)\lambda(X)$ for some polynomial $\omega(X)$. But then $\eta(x_i) = f(x_i) - y_i = \omega(x_i)\lambda(x_i)$ for all $i$, and therefore $\omega(X)$ is an error-evaluator polynomial with respect to $f(X)$, as defined in (69). Furthermore, since $g(X) = \sigma(X)\lambda(X)$ and $\eta(X) = \omega(X)\lambda(X)$, we can rewrite (73) as follows:

$$\psi(X)\,\mathcal{H}(X,Y) = \lambda(X)\bigl(Y\sigma(X) - \omega(X)\bigr)\mathcal{B}(X,Y) \quad (74)$$

Notice that $\eta(x_i) \neq 0$ for all $i \in \mathcal{E}$, and therefore $\omega(x_i) \neq 0$ for all $i \in \mathcal{E}$ as well. Referring to (67), this implies that $\sigma(X)$ and $\omega(X)$ are relatively prime. Consequently, the polynomials $\psi(X)$ and $Y\sigma(X) - \omega(X)$ are also relatively prime. With this, it now follows from (74) that $Y\sigma(X) - \omega(X)$ must be a factor of $\mathcal{H}(X,Y)$, as claimed. ∎

We shall see shortly how the polynomials $\sigma(X)$ and $\omega(X)$ can be extracted from a solution $\mathcal{H}(X,Y)$ to $\mathbf{RIP}_{1,-1}(\mathcal{P}^*, M^*)$. For the time being, let us assume that these polynomial are already known, and explain how decoding should be completed. First, we find the roots of $\sigma(X)$, using either Chien search [15, p. 276] or direct factorization [21]. This reveals the set of error locations $\mathcal{E}$ in (66). Next, we use $\omega(X)$ along with the derivatives of $\sigma(X)$ and $g(X)$ to compute the error values

$$e_i \stackrel{\text{def}}{=} \omega(x_i)\frac{g'(x_i)}{\sigma'(x_i)} \qquad \text{for all } i \in \mathcal{E} \quad (75)$$

Since the set $\mathcal{R} = \{(x_1,y_1),(x_2,y_2),\ldots,(x_k,y_k)\}$ is known to the decoder, this makes it possible to determine the values of the polynomial $f(X)$ at $k$ distinct points, namely

$$f(x_i) = \begin{cases} y_i + e_i & \text{for } i \in \mathcal{E} \\ y_i & \text{for } i \in \mathcal{E}^c \end{cases} \quad (76)$$

Since $\deg f(X) < k$, the entire polynomial $f(X)$ can be recovered from these $k$ values using the same procedure (e.g., Lagrange interpolation) that was used to compute $e(X)$ in (24).

It remains to show how to compute the polynomials $\sigma(X)$ and $\omega(X)$ from $\mathcal{H}(X,Y)$. Note that the ratio $\omega(X)/\sigma(X)$ is a rational function, which can be expanded as the power series

$$\frac{\omega(X)}{\sigma(X)} \stackrel{\text{def}}{=} \gamma_0 + \gamma_1 X + \gamma_2 X^2 + \cdots = \sum_{i=0}^{\infty} \gamma_i X^i \quad (77)$$

provided $\sigma(0) \neq 0$, as we assume[1]. Roth and Ruckenstein [18] view a bivariate polynomial $\mathcal{A}(X,Y)$ as a polynomial in $Y$ with coefficients in the ring $\mathbb{F}_q[X]$, and define a $Y$-root of $\mathcal{A}(X,Y)$ as any element $f(X)$ of $\mathbb{F}_q[X]$ such that $\mathcal{A}(X, f(X))$ is identically zero. They develop an efficient iterative algorithm for computing the $Y$-roots of a given bivariate polynomial. The Roth-Ruckenstein algorithm of [18] takes $d$ iterations to produce all the $Y$-roots of degree at most $d$. Herein, we observe that the Roth-Ruckenstein machinery [18] extends to rational functions. Namely, let us extend the definition of a *$Y$-root of* $\mathcal{A}(X,Y)$ to mean any element $\gamma(X)$ of the field $\mathbb{F}_q(X)$ of rational functions over $\mathbb{F}_q$, such that $\mathcal{A}(X,\gamma(X))$ is identically zero. Then $d$ iterations of the Roth-Ruckenstein algorithm produce the first $d$ coefficients of the power-series expansion of any such $Y$-root. Since $\omega(X)/\sigma(X)$ is a $Y$-root of $\mathcal{H}(X,Y)$ by Theorem 10, we can use the Roth-Ruckenstein algorithm to iteratively generate the coefficients $\gamma_0, \gamma_1, \gamma_2, \ldots$ in (77).

---

[1] We observe that $\sigma(0) = 0$ if and only if $x_i = 0$ for some re-encoding point $(x_i, y_i)$ with $i \in \mathcal{E}$. This situation can be avoided by not including points of type $(0,y)$ in the re-encoding point set $\mathcal{R}$ (in fact, this will happen automatically if the code support set $\mathcal{D}$ is a subset of $\mathbb{F}_q^*$). If one insists on including a $(0,y)$ point in $\mathcal{R}$, we can expand $\omega(X)/\sigma(X)$ in a power series about another point $\alpha$ with $\sigma(\alpha) \neq 0$. In this case, a method similar to the one we describe applies.

How many coefficients do we need? This depends on the degree of $\sigma(X)$, which is equal to the number of errors in the re-encoding positions. Note that $\sigma(X)$ and $\omega(X)$ can be found from the power-series expansion of $\omega(X)/\sigma(X)$ by a Padé approximation procedure, such as the Berlekamp-Massey algorithm. It is well known [15, p. 366] that at most $2t$ coefficients of the power series are required to reconstruct $\omega(X)$ and $\sigma(X)$, where $t = \deg \sigma(X)$. In theory, $\deg \sigma(X) \leqslant k$ in view of (66) and (67), so $2k$ iterations of the Roth-Ruckenstein algorithm always suffice. In practice, we would expect much less than $k$ errors in the $k$ re-encoding positions. Indeed, in soft-decision decoding, we usually choose the re-encoding point set to consist of the $k$ most reliable positions. In this case, the number of errors in these positions will be small with high probability, since most errors occur in the $n-k$ least reliable positions. Consequently, in practice, one would run the Roth-Ruckenstein algorithm for $2\tau$ iterations, where the integer parameter $\tau \leqslant k$ is a pre-determined bound on the number of errors we expect (with high probability) in the $k$ re-encoding positions.

Our results in this section can be summarized in the form of a *reduced factorization algorithm*, presented below.

---

**Reduced Factorization Algorithm**

- **Input:** A solution $\mathcal{H}(X,Y)$ to the reduced interpolation problem $\mathbf{RIP}_{1,-1}(\mathcal{P}^*, M^*)$, an integer parameter $\tau$, and the re-encoding point set $\mathcal{R} = \{(x_1,y_1),(x_2,y_2),\ldots,(x_k,y_k)\}$.
- **Factorization:** Apply the first $2\tau$ iterations of the Roth-Ruckenstein algorithm [18] to $\mathcal{H}(X,Y)$, thereby producing the first $2\tau$ coefficients $\gamma_0, \gamma_1, \ldots, \gamma_{2\tau-1}$ in the power-series expansion of the $Y$-roots of $\mathcal{H}(X,Y)$. There are at most $r$ such $Y$-roots, where $r = \deg_{0,1}\mathcal{H}(X,Y)$.
- **Padé Approximation:** Treat each sequence $\gamma_0, \gamma_1, \ldots, \gamma_{2\tau-1}$ produced in the previous step as a "syndrome" sequence; then compute the coefficients $\sigma_0, \sigma_1, \ldots, \sigma_t$ of the shortest linear feedback shift-register that generates this syndrome sequence (Berlekamp-Massey algorithm). Next, find the coefficients $\omega_0, \omega_1, \ldots, \omega_t$ of the error-evaluator polynomial $\omega(X)$, e.g., using the convolution:

$$\omega_i := \sigma_0 \gamma_i + \sigma_1 \gamma_{i-1} + \cdots + \sigma_{i-1}\gamma_1 + \sigma_i \gamma_0 \quad (78)$$

- **Root Finding:** Find all the roots of the error-locator polynomial $\sigma(X) = \sigma_0 + \sigma_1 X + \cdots + \sigma_t X^t$, and for each such root compute the corresponding error value using (75).
- **Corrected Re-Encoding:** Set the $k$ corrected re-encoding values $f(x_1), f(x_2), \ldots, f(x_k)$ as in (76), and find the unique polynomial $f(X)$ of degree $< k$ that agrees with these values (using, for example, Lagrange interpolation).
- **Output:** Return the list of all such polynomials $f(X)$.

---

Not all the $Y$-roots produced by the $2\tau$ iterations of the Roth-Ruckenstein algorithm necessarily correspond to a valid $\sigma(X)$ and $\omega(X)$ pair. Just as in conventional Berlekamp-Massey decoding, we expect the error-locator polynomial $\sigma(X)$ and the error-evaluator polynomial $\omega(X)$ to satisfy certain conditions. If these conditions are violated, the corresponding $Y$-root can be rejected. For example, a $Y$-root sequence $\gamma_0, \gamma_1, \ldots, \gamma_{2\tau-1}$ can be safely rejected if any of the following occurs:

- **a)** the degree $t$ of $\sigma(X)$ is strictly greater than $\tau$;
- **b)** the convolution (78) produces nonzero values for $i > t$;
- **c)** the polynomial $\sigma(X)$ has less than $t$ distinct roots in $\mathbb{F}_q$;
- **d)** an error value $e_i = 0$ is computed according to (75).

We leave the proof of the underlying properties of $\sigma(X)$ and $\omega(X)$ as an exercise for the reader. In practice, all these conditions are used to detect false $Y$-roots in the reduced factorization algorithm. As a result, with very high probability, only *one* polynomial $f(X)$ is returned by the algorithm.

**Example 2d.** Consider again the situation in Example 2. Recall that the solution $\mathcal{Q}(X,Y)$ to the original interpolation problem, found by the Koetter algorithm, factors as follows:

$$\mathcal{Q}(X,Y) = (\alpha^3 + X)\big(Y - (\alpha^6 + \alpha^2 X)\big)\big(Y - (\alpha^5 + \alpha^6 X)\big)$$

The re-encoding point set in Example 2a consists of the points $(x_1,y_1) = (\alpha, \alpha^4)$ and $(x_2,y_2) = (\alpha^2, \alpha^6)$, and the resulting re-encoding polynomial is $e(X) = \alpha^5 + \alpha^6 X$. Solving the reduced interpolation problem produces the polynomial

$$\mathcal{H}(X,Y) = (\alpha^3 + X)Y + (\alpha^5 + \alpha^5 X + X^2)Y^2$$

as shown in Example 2c (cf. (63) and the last row of Table III). Observe that this polynomial factors as follows:

$$\mathcal{H}(X,Y) = (\alpha^5 + \alpha^2 X)\, Y\, \big(Y(1 + \alpha^5 X) - \alpha^5\big) \quad (79)$$

Applying the Roth-Ruckenstein algorithm [18] to $\mathcal{H}(X,Y)$ produces two syndrome sequences. The first sequence, which corresponds to the factor $Y$ in (79), is identically zero. The shortest feedback shift-register that generates the all-zero sequence is $\sigma(X) = 1$, and the corresponding error-evaluator polynomial is $\omega(X) = 0$. This indicates the absence of errors in the re-encoding positions, and the resulting message polynomial is $f_1(X) = e(X) = \alpha^5 + \alpha^6 X$. Note that $Y - f_1(X)$ is, indeed, a factor of $\mathcal{Q}(X,Y)$. The second syndrome sequence, corresponding to the factor $Y(1 + \alpha^5 X) - \alpha^5$ in (79), is given by

$$\gamma_0, \gamma_1, \gamma_2, \ldots = \alpha^5, \alpha^3, \alpha, \alpha^6, \alpha^4, \alpha^2, 1, \alpha^5, \ldots \quad (80)$$

Applying the Berlekamp-Massey algorithm to the syndrome sequence (80), we find that it is generated by $\sigma(X) = 1 + \alpha^5 X$. The convolution (78) then produces $\omega_0 = \alpha^5$ and $\omega_1 = 0$, so that $\omega(X) = \alpha^5$. The error-locator polynomial has a single root at $x_2 = \alpha^2$. Thus the set of error locations is $\mathcal{E} = \{2\}$, and the single error value is given by

$$e_2 = \omega(x_2)\frac{g'(x_2)}{\sigma'(x_2)} = \alpha^5\frac{\alpha^4}{\alpha^5} = \alpha^4$$

In accordance with (76), we now set $f_2(x_2) = y_2 + e_2 = \alpha^3$ and $f_2(x_1) = y_1 = \alpha^4$. The unique polynomial of degree $< 2$ that has these values is $f_2(X) = \alpha^6 + \alpha^2 X$. This polynomial, indeed, corresponds to the other factor of $\mathcal{Q}(X,Y)$, and to the transmitted codeword (cf. Example 2a). □

## VI. Conclusions

A complete proof of the applicability of the re-encoding and coordinate transformation method to the bivariate polynomial interpolation process in algebraic list-decoding of Reed-Solomon codes is presented. A detailed example is given to illustrate the entire re-encoding and coordinate transformation process. In addition, it is shown how the factorization procedure can be modified to accommodate the reduced interpolation problem. Thereby, factorization complexity is also decreased significantly since the required number of iterations of the Roth-Ruckenstein factorization algorithm is reduced from $k$ to $2\tau$, where $\tau$ is a small number (we found that, in practice, $\tau \leqslant 6$ often suffices, even for long high-rate Reed-Solomon codes).

| $(x,y)$ | $m_{x,y}$ | Gröbner Basis Polynomials |
|---|---|---|
| $(\alpha, \alpha^4)$ | 2 | $\mathcal{G}_0(X,Y) = \alpha + X$ <br> $\mathcal{G}_1(X,Y) = \alpha^4 + Y$ <br> $\mathcal{G}_2(X,Y) = \alpha + Y^2$ <br> $\mathcal{G}_3(X,Y) = \alpha^5 + Y^3$ |
| | | $\mathcal{G}_1(X,Y) = \alpha^4 + Y$ <br> $\mathcal{G}_0(X,Y) = \alpha^2 + X^2$ <br> $\mathcal{G}_2(X,Y) = \alpha + Y^2$ <br> $\mathcal{G}_3(X,Y) = \alpha^5 + Y^3$ |
| | | $\mathcal{G}_0(X,Y) = \alpha^2 + X^2$ <br> $\mathcal{G}_1(X,Y) = (\alpha^5 + \alpha^4 X) + (\alpha + X)Y$ <br> $\mathcal{G}_2(X,Y) = \alpha + Y^2$ <br> $\mathcal{G}_3(X,Y) = \alpha Y + Y^3$ |
| $(\alpha^2, \alpha^6)$ | 1 | $\mathcal{G}_1(X,Y) = (\alpha^6 + \alpha^4 X + \alpha^6 X^2) + (\alpha + X)Y$ <br> $\mathcal{G}_2(X,Y) = (\alpha^3 + \alpha^5 X^2) + Y^2$ <br> $\mathcal{G}_0(X,Y) = \alpha^4 + \alpha^2 X + \alpha^2 X^2 + X^3$ <br> $\mathcal{G}_3(X,Y) = (\alpha^6 + \alpha^4 X^2) + \alpha Y + Y^3$ |
| $(\alpha^2, \alpha^3)$ | 1 | $\mathcal{G}_2(X,Y) = (\alpha^4 + \alpha^4 X + \alpha X^2) + (\alpha + X)Y + Y^2$ <br> $\mathcal{G}_0(X,Y) = \alpha^4 + \alpha^2 X + \alpha^2 X^2 + X^3$ <br> $\mathcal{G}_1(X,Y) = (\alpha + \alpha^2 X^2 + \alpha^6 X^3) + (\alpha^3 + \alpha^4 X + X^2)Y$ <br> $\mathcal{G}_3(X,Y) = (\alpha^3 + \alpha^2 X) + (\alpha^5 + \alpha^5 X)Y + Y^3$ |
| $(\alpha^3, 1)$ | 1 | $\mathcal{G}_0(X,Y) = (1 + \alpha^3 X + X^3) + (\alpha^2 + \alpha X)Y + \alpha Y^2$ <br> $\mathcal{G}_1(X,Y) = (\alpha^5 + \alpha^6 X + \alpha^5 X^2 + \alpha^6 X^3) + (\alpha X + X^2)Y + \alpha^2 Y^2$ <br> $\mathcal{G}_2(X,Y) = (1 + \alpha^5 X + \alpha X^3) + (\alpha^4 + X + X^2)Y + (\alpha^3 + X)Y^2$ <br> $\mathcal{G}_3(X,Y) = (\alpha^3 + \alpha^2 X) + (\alpha^5 + \alpha^5 X)Y + Y^3$ |
| $(\alpha^3, \alpha)$ | 1 | $\mathcal{G}_1(X,Y) = (\alpha^4 + \alpha^4 X + \alpha^5 X^2 + \alpha^2 X^3) + (\alpha^2 + X^2)Y + \alpha^4 Y^2$ <br> $\mathcal{G}_2(X,Y) = (1 + \alpha^5 X + \alpha X^3) + (\alpha^4 + X + X^2)Y + (\alpha^3 + X)Y^2$ <br> $\mathcal{G}_3(X,Y) = (\alpha^4 + \alpha^6 X^3) + (\alpha^6 + \alpha^4 X)Y + Y^2 + Y^3$ <br> $\mathcal{G}_0(X,Y) = (\alpha^3 + \alpha^2 X + \alpha^3 X^2 + \alpha^3 X^3 + X^4) + (\alpha^5 + \alpha X + \alpha X^2)Y + (\alpha^4 + \alpha X)Y^2$ |
| $(1, \alpha)$ | 1 | $\mathcal{G}_2(X,Y) = (1 + \alpha^5 X + \alpha X^3) + (\alpha^4 + X + X^2)Y + (\alpha^3 + X)Y^2$ <br> $\mathcal{G}_3(X,Y) = (\alpha^5 + X + \alpha X^2 + \alpha X^3) + (\alpha + \alpha^4 X + \alpha^3 X^2)Y + Y^3$ <br> $\mathcal{G}_0(X,Y) = (\alpha^2 + \alpha^3 X + \alpha^4 X^2 + X^4) + (\alpha^2 + \alpha X)Y + (1 + \alpha X)Y^2$ <br> $\mathcal{G}_1(X,Y) = (\alpha^4 + X^2 + \alpha^3 X^3 + \alpha^2 X^4) + (\alpha^2 + \alpha^2 X + X^2 + X^3)Y + (\alpha^4 + \alpha^4 X)Y^2$ |
| $(1, 1)$ | 1 | $\mathcal{G}_2(X,Y) = (1 + \alpha^5 X + \alpha X^3) + (\alpha^4 + X + X^2)Y + (\alpha^3 + X)Y^2$ <br> $\mathcal{G}_3(X,Y) = (\alpha^5 + X + \alpha X^2 + \alpha X^3) + (\alpha + \alpha^4 X + \alpha^3 X^2)Y + Y^3$ <br> $\mathcal{G}_1(X,Y) = (\alpha^4 + X^2 + \alpha^3 X^3 + \alpha^2 X^4) + (\alpha^2 + \alpha^2 X + X^2 + X^3)Y + (\alpha^4 + \alpha^4 X)Y^2$ <br> $\mathcal{G}_0(X,Y) = (\alpha^2 + \alpha^5 X + \alpha^6 X^2 + \alpha^4 X^3 + X^4 + X^5) + (\alpha^2 + \alpha^4 X + \alpha X^2)Y + (1 + \alpha^3 X + \alpha X^2)Y^2$ |

TABLE I

ITERATIONS OF THE KOETTER ALGORITHM FOR THE ORIGINAL INTERPOLATION PROBLEM IN EXAMPLE 2a

| $(x,y)$ | $m_{x,y}$ | Gröbner Basis Polynomials |
|---|---|---|
| $(\alpha,0)$ | 2 | $\mathcal{G}_0(X,Y) = \alpha + X$ <br> $\mathcal{G}_1(X,Y) = Y$ <br> $\mathcal{G}_2(X,Y) = Y^2$ <br> $\mathcal{G}_3(X,Y) = Y^3$ |
| | | $\mathcal{G}_0(X,Y) = \alpha + X$ <br> $\mathcal{G}_1(X,Y) = (\alpha + X)Y$ <br> $\mathcal{G}_2(X,Y) = Y^2$ <br> $\mathcal{G}_3(X,Y) = Y^3$ |
| | | $\mathcal{G}_0(X,Y) = \alpha^2 + X^2$ <br> $\mathcal{G}_1(X,Y) = (\alpha + X)Y$ <br> $\mathcal{G}_2(X,Y) = Y^2$ <br> $\mathcal{G}_3(X,Y) = Y^3$ |
| $(\alpha^2,0)$ | 1 | $\mathcal{G}_0(X,Y) = \alpha^4 + \alpha^2 X + \alpha^2 X^2 + X^3$ <br> $\mathcal{G}_1(X,Y) = (\alpha + X)Y$ <br> $\mathcal{G}_2(X,Y) = Y^2$ <br> $\mathcal{G}_3(X,Y) = Y^3$ |
| $(\alpha^3,\alpha)$ | 1 | $\mathcal{G}_2(X,Y) = (\alpha^2 + \alpha X)Y + Y^2$ <br> $\mathcal{G}_0(X,Y) = (\alpha^4 + \alpha^2 X + \alpha^2 X^2 + X^3) + (\alpha^5 + \alpha^4 X)Y$ <br> $\mathcal{G}_1(X,Y) = (\alpha^4 + X + X^2)Y$ <br> $\mathcal{G}_3(X,Y) = (\alpha^3 + \alpha^2 X)Y + Y^3$ |
| $(\alpha^3,1)$ | 1 | $\mathcal{G}_0(X,Y) = (\alpha^4 + \alpha^2 X + \alpha^2 X^2 + X^3) + (\alpha + X)Y + \alpha^4 Y^2$ <br> $\mathcal{G}_1(X,Y) = (\alpha^4 + X + X^2)Y$ <br> $\mathcal{G}_2(X,Y) = (\alpha^5 + \alpha X + \alpha X^2)Y + (\alpha^3 + X)Y^2$ <br> $\mathcal{G}_3(X,Y) = (\alpha^2 + \alpha X)Y + \alpha^3 Y^2 + Y^3$ |
| $(1,0)$ | 1 | $\mathcal{G}_1(X,Y) = (\alpha^4 + X + X^2)Y$ <br> $\mathcal{G}_2(X,Y) = (\alpha^5 + \alpha X + \alpha X^2)Y + (\alpha^3 + X)Y^2$ <br> $\mathcal{G}_3(X,Y) = (\alpha^2 + \alpha X)Y + \alpha^3 Y^2 + Y^3$ <br> $\mathcal{G}_0(X,Y) = (\alpha^4 + \alpha X + \alpha^6 X^3 + X^4) + (\alpha + \alpha^3 X + X^2)Y + (\alpha^4 + \alpha^4 X)Y^2$ |
| $(1,\alpha^3)$ | 1 | $\mathcal{G}_2(X,Y) = (\alpha^4 + X + X^2)Y + (\alpha^3 + X)Y^2$ <br> $\mathcal{G}_3(X,Y) = (\alpha + \alpha^3 X + X^2)Y + \alpha^3 Y^2 + Y^3$ <br> $\mathcal{G}_0(X,Y) = (\alpha^4 + \alpha X + \alpha^6 X^3 + X^4) + (\alpha + \alpha^3 X + X^2)Y + (\alpha^4 + \alpha^4 X)Y^2$ <br> $\mathcal{G}_1(X,Y) = (\alpha^4 + \alpha^5 X + X^3)Y$ |
| $(\alpha^2,\alpha^4)$ | 1 | $\mathcal{G}_2(X,Y) = (\alpha^4 + X + X^2)Y + (\alpha^3 + X)Y^2$ <br> $\mathcal{G}_3(X,Y) = (\alpha + \alpha^3 X + X^2)Y + \alpha^3 Y^2 + Y^3$ <br> $\mathcal{G}_1(X,Y) = (\alpha^4 + \alpha X + \alpha^6 X^3 + X^4) + (\alpha^2 + \alpha^2 X + X^2 + X^3)Y + (\alpha^4 + \alpha^4 X)Y^2$ <br> $\mathcal{G}_0(X,Y) = (\alpha^6 + \alpha^6 X + \alpha X^2 + \alpha X^3 + X^4 + X^5) + (\alpha^3 + \alpha^6 X + \alpha^5 X^2 + X^3)Y + (\alpha^6 + \alpha^3 X + \alpha^4 X^2)Y^2$ |

TABLE II

Iterations of the Koetter Algorithm for the Shifted Interpolation Problem in Example 2a

| $(x,z)$ | $m_{x,z}$ | Gröbner Basis Polynomials |
|---|---|---|
| $(\alpha^3, \alpha^3)$ | 1 | $\mathcal{G}_2(X,Y) = \alpha Y + (\alpha^2 + X)Y^2$ <br> $\mathcal{G}_0(X,Y) = 1 + \alpha^4 Y$ <br> $\mathcal{G}_1(X,Y) = (\alpha^3 + X)Y$ <br> $\mathcal{G}_3(X,Y) = \alpha^2 Y + (\alpha^5 + \alpha^4 X + \alpha X^2 + X^3)Y^3$ |
| $(\alpha^3, \alpha^2)$ | 1 | $\mathcal{G}_0(X,Y) = 1 + Y + (\alpha^6 + \alpha^4 X)Y^2$ <br> $\mathcal{G}_1(X,Y) = (\alpha^3 + X)Y$ <br> $\mathcal{G}_2(X,Y) = (\alpha^4 + \alpha X)Y + (\alpha^5 + \alpha^5 X + X^2)Y^2$ <br> $\mathcal{G}_3(X,Y) = \alpha Y + (\alpha^5 + \alpha^3 X)Y^2 + (\alpha^5 + \alpha^4 X + \alpha X^2 + X^3)Y^3$ |
| $(1, 0)$ | 1 | $\mathcal{G}_1(X,Y) = (\alpha^3 + X)Y$ <br> $\mathcal{G}_2(X,Y) = (\alpha^4 + \alpha X)Y + (\alpha^5 + \alpha^5 X + X^2)Y^2$ <br> $\mathcal{G}_3(X,Y) = \alpha Y + (\alpha^5 + \alpha^3 X)Y^2 + (\alpha^5 + \alpha^4 X + \alpha X^2 + X^3)Y^3$ <br> $\mathcal{G}_0(X,Y) = (1 + X) + (1 + X)Y + (\alpha^6 + \alpha^3 X + \alpha^4 X^2)Y^2$ |
| $(1, \alpha)$ | 1 | $\mathcal{G}_2(X,Y) = (\alpha^3 + X)Y + (\alpha^5 + \alpha^5 X + X^2)Y^2$ <br> $\mathcal{G}_3(X,Y) = (1 + X)Y + (\alpha^5 + \alpha^3 X)Y^2 + (\alpha^5 + \alpha^4 X + \alpha X^2 + X^3)Y^3$ <br> $\mathcal{G}_0(X,Y) = (1 + X) + (1 + X)Y + (\alpha^6 + \alpha^3 X + \alpha^4 X^2)Y^2$ <br> $\mathcal{G}_1(X,Y) = (\alpha^3 + \alpha X + X^2)Y$ |
| $(\alpha^2, 1)$ | 1 | $\mathcal{G}_2(X,Y) = (\alpha^3 + X)Y + (\alpha^5 + \alpha^5 X + X^2)Y^2$ <br> $\mathcal{G}_3(X,Y) = (1 + X)Y + (\alpha^5 + \alpha^3 X)Y^2 + (\alpha^5 + \alpha^4 X + \alpha X^2 + X^3)Y^3$ <br> $\mathcal{G}_1(X,Y) = (1 + X) + (\alpha + \alpha^3 X + X^2)Y + (\alpha^6 + \alpha^3 X + \alpha^4 X^2)Y^2$ <br> $\mathcal{G}_0(X,Y) = (\alpha^2 + \alpha^6 X + X^2) + (\alpha^2 + \alpha^6 X + X^2)Y + (\alpha + \alpha X + \alpha^4 X^2 + \alpha^4 X^3)Y^2$ |

TABLE III

ITERATIONS OF THE KOETTER ALGORITHM FOR THE REDUCED INTERPOLATION PROBLEM IN EXAMPLE 2c